\documentclass[11pt,prd,aps,showpacs,eqsecnum,floatfix,nofootinbib,preprint,tightenlines]{revtex4}
\usepackage{graphicx}				% Use pdf, png, jpg, or eps§ with pdflatex; use eps in DVI mode
\usepackage{latexsym}
\usepackage{epsf}
\usepackage{graphicx}
\usepackage{slashed}
\usepackage{amsmath}
\usepackage{amssymb}
\usepackage{color}
\usepackage{colordvi}
\usepackage{multirow}
\newcommand{\be}{\begin{equation}}
\newcommand{\ee}{\end{equation}}
\newcommand{\bea}{\begin{eqnarray}}
\newcommand{\eea}{\end{eqnarray}}
\newcommand{\ba}{\begin{array}}
\newcommand{\ea}{\end{array}}
\newcommand{\nn}{\nonumber}

\newcommand{\pref}[1]{(\ref{#1})}
\newcommand{\np}{N\pi}

\newcommand{\Epipmq}{E_{\pi, \vec{p}-\vec{q}}}
\newcommand{\Epippq}{E_{\pi, \vec{p}+\vec{q}}}

\newcommand{\ENq}{E_{N,\vec{q}}}

\newcommand{\Epip}{E_{\pi, \vec{p}}}
\newcommand{\Epir}{E_{\pi,{\vec{r}}}}
\newcommand{\Epis}{E_{\pi,{\vec{s}}}}

\newcommand{\GE}{G_{\rm E}}
\newcommand{\GM}{G_{\rm M}}
\newcommand{\GEf}{G_{{\rm E},{4}}}
\newcommand{\GEi}{G_{{\rm E},i}}
\newcommand{\eV}[1]{\epsilon^{\rm mid}_{{\rm E},{#1}}}
\newcommand{\eM}{\epsilon^{\rm mid}_{\rm M}}
\newcommand{\GEmid}{G^{\rm mid}_{\rm E,4}}
\newcommand{\GMmid}{G^{\rm mid}_{\rm M}}

\newcommand{\psq}{|\vec{p}|^2}
\newcommand{\qsq}{|\vec{q}|^2}
\newcommand{\pq}{\vec{p} \cdot \vec{q}}

\begin{document}
\renewcommand{\thefootnote}{$*$}

\preprint{HU-EP-21/04}

\title{$N\pi$-state contamination in lattice calculations of the nucleon electromagnetic form factors}

\author{Oliver B\"ar$^{a}$ and Haris \v{C}oli\'{c}$^{a}$} 
\affiliation{$^a$Institut f\"ur Physik,
\\Humboldt Universit\"at zu Berlin,
\\12489 Berlin, Germany\\}

\begin{abstract}
The nucleon-pion-state contribution to QCD two-point and three-point functions relevant for lattice calculations of the nucleon electromagnetic form factors are studied in chiral perturbation theory. 
To leading order the results depend on a few experimentally known low-energy constants only, and the nucleon-pion-state contribution to the form factors can be estimated. The nucleon-pion-state contribution to the electric form factor $G_{\rm E}(Q^2)$ is at the +5 percent level for a source-sink separation of 2 fm, and it increases with increasing momentum transfer $Q^2$. For the magnetic form factor  the nucleon-pion-state contribution leads to an  underestimation of $G_{\rm M}(Q^2)$ by about 5 percent that decreases with increasing $Q^2$. 
For smaller source-sink separations that are accessible in present-day lattice simulations the impact is larger, although  the ChPT results may not be applicable for such small  time separations. Still, a comparison with lattice data at $t\approx 1.6$ fm works reasonably well.
\end{abstract}

\pacs{11.15.Ha, 12.39.Fe, 12.38.Gc}
\maketitle

\renewcommand{\thefootnote}{\arabic{footnote}} \setcounter{footnote}{0}

\newpage
%========================
\section{Introduction}\label{Intro} 
%========================

The internal structure of nucleons is conveniently described by electromagnetic form factors. 
Experimentally these quantities are accessible by elastic electron-nucleon scattering experiments. Such experiments have a long history, going back to the Hofstadter experiments in the 1950s to recent experiments at Mainz, JLab and MIT-Bates. For a review of these experimental efforts see Ref.\ \cite{Punjabi:2015bba}.
Theoretically, the form factors can be calculated numerically in lattice QCD simulations. The computational techniques are well-established, but for the numerical results to have phenomenological impact reliable results with controlled errors are needed. 

For years the chiral extrapolation  has been among the dominant sources for the systematic error in lattice results. Today, increased computer power and improved simulation algorithms allow physical point simulations with the quark masses set to their physical values, eliminating the need for a chiral extrapolation and the associated systemantic uncertainty. Instead, the excited-state contamination is widely accepted to cause the dominant systematic error in many lattice QCD results. Physical point simulations in particular are afflicted with an excited state contamination due to multi-particle states involving light pions. For a recent review of the excited-state impact on nucleon structure observables see Ref.\ \cite{Ottnad:2020qbw}.

In a series of papers \cite{Bar:2018xyi,Bar:2019gfx,Bar:2019igf}, chiral perturbation theory (ChPT) \cite{Weinberg:1978kz,Gasser:1983yg,Gasser:1984gg}, the low-energy effective theory of QCD, was employed to study the  excited state contamination due to 2-particle nucleon-pion ($N\pi$) states in the axial and pseudoscalar form factors of the nucleon. The leading order (LO) results were found to describe surprisingly well various discrepancies between the lattice plateau estimates and the phenomenologically expected results for the form factors. In particular, ChPT provides an explanation for the so-called PCAC puzzle \cite{Rajan:2017lxk,Bali:2018qus}: the apparent violation of the generalised Goldberger-Treiman relation between the axial and pseudoscalar form factors is caused by the contribution of a low-energetic $N\pi$ state in the induced pseudoscalar form factor.
This conclusion is supported by the ChPT analysis \cite{Bar:2019igf} of the projection method proposed in Ref.\ \cite{Bali:2018qus} to solve the PCAC puzzle. Even though this method is found to be insufficient, soon thereafter an alternative strategy to deal with the $N\pi$ contamination was proposed and studied with promising results \cite{Jang:2019vkm,Bali:2019yiy}.

In this paper we report our ChPT results for the $N\pi$ contamination in the electromagnetic nucleon form factors. The calculational setup is essentially as in Ref.\ \cite{Bar:2018xyi}, with the axial vector current replaced by the vector current. This replacement leads to many changes in the details and the final results, the most notable one being the absence of a dominant low-energetic $N\pi$-state contribution as in the induced pseudoscalar form factor. Physically it stems from the ability of the axial vector current to emit (absorb) a pion that is absorbed (emitted) at the sink (source) of the three-point (3-pt) that needs to be computed to obtain the form factors. This is not allowed for the vector current, two pions instead of one are needed for the analogous process, and the resulting 3-particle $N\pi\pi$ contamination is expected to be substantially smaller.\footnote{The 3-particle $N\pi\pi$ contamination in the nucleon 2-pt function was computed in Ref.\ \cite{Bar:2018wco} and found to be negligible compared to the 2-particle $N\pi$ contamination.} 

The main results of this paper can be summarised as follows. In the common plateau and midpoint estimates the $N\pi$ state contamination leads to an overestimation of the electric form factor, and the misestimation increases with increasing momentum transfer $Q^2$. In contrast, the magnetic form factor is underestimated, and the misestimation gets larger for smaller $Q^2$.
How big this effect is depends on the source-sink separation $t$ assumed for the vector current 3pt-function. For $t=2$ fm the misestimation is at the $\pm$ 5\% level for lattice simulations with physical pion masses. 
The impact increases for the smaller values $t\lesssim 1.5$ fm that are accessible in present-day simulations. Applying the ChPT results to such small source-sink separations is problematic, for the correlation functions are not expected to be dominated by pion physics. We nevertheless find good agreement when we compare the ChPT predictions with recent lattice results in \cite{Ishikawa:2018rew,Alexandrou:2018sjm}. Moreover, various observations in \cite{Jang:2019jkn} about the excited-state contamination in the electric form factor obtained with the spatial components of the vector current 3pt-function are qualitatively explained by ChPT.

The calculational setup employed here is essentially the same as in Ref.\ \cite{Bar:2018xyi,Bar:2016uoj,Bar:2016jof} and is only briefly reviewed in the following. The methodology for studying the excited state contamination using ChPT goes back to  Refs.\cite{Tiburzi:2009zp,Bar:2012ce}. It is also reviewed in \cite{Bar:2017kxh,Bar:2017gqh} to which the reader is referred to for more details.

%========================
\section{The electromagnetic form factors}\label{sec:ff} 
%========================

\subsection{The electromagnetic form factors of the nucleon}
We start with summarising some basic definitions to settle our notation.
The matrix element of the electromagnetic current 
\begin{equation}
V^{\mu}_{\rm em} = \frac{2}{3} \overline{u}\gamma^{\mu}u - \frac{1}{3} \overline{d}\gamma^{\mu}d\,\ldots 
\end{equation}
between single-nucleon states can be decomposed in terms of the Dirac and Pauli form factors $F^N_1$ and $F^N_2$,
\begin{equation}\label{DefFF1}
\langle N(p',s')|V^{\mu}_{\rm em}(0)|N(p,s)\rangle = \bar{u}(p',s')\left(\gamma^{\mu}F^N_{1}(q^2) + i \frac{\sigma^{\mu\nu} q_{\nu}}{2M_N}F^N_2(q^2)\right)u(p,s)\,.
\end{equation}
$N=p$ or $n$ refers to either the proton or neutron as the nucleon, and $u(p,s)$ is a Dirac spinor with momentum $p$ and spin $s$. $\sigma^{\mu\nu}=\frac{i}{2}[\gamma^{\mu},\gamma^{\nu}]$ is the standard Clifford algebra element formed from the Dirac matrices $\gamma^{\mu}$. $M_N$ denotes the nucleon mass, and the four-momentum transfer $q=p'-p$ is given by 
\begin{equation}
q^2=-Q^2=(E_{N,\vec{p}^{\,\prime}} -  E_{N,\vec{p}})^2 - (\vec{p}^{\prime}-\vec{p})^2\,,
\end{equation}
with $E_{N,\vec{p}}=\sqrt{|\vec{p}|^{\,2} +M_N^2}$ denoting the  energy of a nucleon with spatial momentum $\vec{p}$.

Throughout this paper we assume isospin symmetry with degenerate up and down quark masses. In that case one finds the relation  
 \cite{Alexandrou:2017hac}
\begin{eqnarray}\label{VCME}
\langle p|  \overline{u}\gamma^{\mu}u - \overline{d}\gamma^{\mu}d | p\rangle 
&=& \langle p|V^{\mu}_{\rm em} | p\rangle - \langle n|V^{\mu}_{\rm em} | n\rangle\,,
\end{eqnarray}
where we suppress the dependency on the momenta and spins of proton and neutron. The matrix element on the left hand side contains the flavour non-singlet vector current.  Performing the form factor decomposition for this matrix element we obtain the same result as in \pref{DefFF1} but with the the non-singlet Dirac and Pauli form factors $F^{u-d}_{1,2}$. For brevity we drop the index $u-d$ in the following, thus we find
\begin{equation}
F_1(q^2) = F^p_1(q^2) - F^n_1(q^2),\qquad F_2(q^2) = F^p_2(q^2) - F^n_2(q^2),
\end{equation}
for the non-singlet form factors.

In practice it is convenient to use linear combinations of these form factors,
\begin{eqnarray}
\GE(q^2)=F_1(q^2) + \frac{q^2}{4M_N^2}F_2(q^2)\,,\label{SachsFFGE}\\
\GM(q^2)=F_1(q^2)+F_2(q^2)\,,\label{SachsFFGM}
\end{eqnarray}
with the electric and magnetic (Sachs) form factors $\GE$ and $\GM$. These can be determined from electron-nucleon scattering data \cite{Punjabi:2015bba}. In addition, the slope of the form factors at vanishing momentum transfer defines the charge radii squared, 
\begin{eqnarray}\label{DefRadii}
\overline{r}^2_{\rm X}\equiv \langle r_{\rm X}^2\rangle = -6 \frac{d}{d Q^2} \left(\frac{G_{\rm X}(Q^2)}{G_{\rm X}(0)}\right)\bigg|_{Q^2=0}\,,
\end{eqnarray}
with ${\rm X=E,M}$.

\subsection{Lattice calculation of the form factors}
The electromagnetic form factors are accessible in lattice QCD simulations with spacelike momentum transfer $q^2=-Q^2 <0$. 
The standard procedure is based on evaluating various Euclidean 2- and 3-point (pt) functions. Explicitly, the nucleon 2-pt function is given by\footnote{We continue to use the continuum formulation for all expressions even if we explicitly refer to correlation functions measured on a discrete space-time lattice.}
\begin{equation}\label{Def:2ptfunc}
C_2(\vec{p},t)= \int d^3x \,e^{i\vec{p}\vec{x}}\, \Gamma_{\beta\alpha}\langle N_{\alpha}(\vec{x},t)\overline{N}_{\beta}(0,0)\rangle \,.
\end{equation}
$N,\overline{N}$ denote nucleon interpolating fields placed at sink (Euclidean time $t$) and source ($t=0$). Although arbitrary to a large extent we assume them to be given by the standard 3-quark operators (either pointlike or smeared) that have been mapped to ChPT \cite{Bar:2015zwa}. The matrix $\Gamma$ acts on spinor space and is given by
\begin{equation}\label{DefProjGamma}
\Gamma=\frac{1+\gamma_4}{4}(1+i \gamma_5\gamma_3)\,.
\end{equation}
This definition corresponds to the one employed in \cite{Alexandrou:2017hac} by the ETM collaboration, but differs by a factor 1/2 from the one used in \cite{Capitani:2017qpc}, for example. 
This difference, however, is irrelevant since the form factors are obtained from ratios of correlation functions where the different normalisation drops out.

The form factors depend on the momentum transfer $Q^2$ only. Therefore, the nucleon 3-pt function can be  computed with some simplifying kinematics: The nucleon at the sink is chosen to be at rest, i.e.\ $\vec{p}^{\,\prime}=0$, which implies $\vec{q}=-\vec{p}$ and 
\begin{equation}\label{DefQsqr}
Q^2=\vec{q}^{\,2} - (M_N-E_{N,\vec{q}})^2,
\end{equation}
for the momentum transfer. According to \pref{VCME}  we choose the third isospin component of the vector current, $a=3$, in terms of the standard basis with the familiar Pauli matrices. Therefore, the 3-pt function we consider reads
\begin{equation}\label{Def3ptfunc}
C_{3,{\mu}}(\vec{q},t,t')\equiv C_{3,V^3_{\mu}}(\vec{q},t,t')=\int d^3x\int d^3y \,e^{i\vec{q}\vec{y}}\, \Gamma_{\beta\alpha}\langle N_{\alpha}(\vec{x},t) V_{\mu}^3(\vec{y},t')\overline{N}_{\beta}(0,0)\rangle\,,
\end{equation}
with the Euclidean time $t'$ denoting the operator insertion time.

With the 2-pt and 3-pt functions we form the generalised ratio
\begin{equation}\label{DefRatio}
R_{\mu}(\vec{q},t,t') =\frac{C_{3,{\mu}}(\vec{q},t,t')}{C_2(0,t)}\sqrt{\frac{C_2(\vec{q},t-t')}{C_2(0,t-t')}\frac{C_2(0,t)}{C_2(\vec{q},t)}\frac{C_2(0,t')}{C_2(\vec{q},t')}} \ .
\end{equation}
By construction this ratio converges to constant asymptotic values,
\begin{equation}\label{Pimu}
R_{\mu}(\vec{q},t,t')  \longrightarrow \Pi_{{\mu}}(\vec{q}),
\end{equation}
in the limit $t,t', t-t'\rightarrow \infty$,
and these are trivially related to the electromagnetic form factors \cite{Capitani:2015sba,Jang:2019jkn}:
\begin{eqnarray}
{\rm Re}\,\Pi_{4}(\vec{q}) &=&  \sqrt{\frac{E_{N,\vec{q}}+M_N}{2E_{N,\vec{q}}}} \, \GE(Q^2)\,,\label{AsympValuePi4}\\
{\rm Re}\,\Pi_{i}(\vec{q}) &=& \epsilon_{ij3}  q_j\frac{1}{\sqrt{2 E_{N,\vec{q}}(E_{N,\vec{q}}+M_N)}}\, \GM(Q^2) \,,\label{AsympValueRePik}\\
{\rm Im}\,\Pi_{i}(\vec{q}) &=& q_j\frac{1}{\sqrt{2 E_{N,\vec{q}}(E_{N,\vec{q}}+M_N)}}\, \GE(Q^2) \,.\label{AsympValueImPik}
\end{eqnarray}
Thus, the form factor are obtained from the $\Pi_{{\mu}}(\vec{q})$ by multiplication with some simple kinematical factors involving the nucleon's energy and spatial momentum.

\subsection{Vector current conservation}
For degenerate up and down quark masses the vector current is conserved,
\begin{equation}\label{VCC}
\partial_{\mu}V^a_{\mu}(x) = 0\,,
\end{equation}
and we find three conserved charges $Q^a$, $a=1,2,3$. 
With our conventions the conserved charges are $Q^a=1$.

Current conservation implies a Ward identity for the correlation functions we have introduced in the last subsection. The 3-pt function \pref{Def3ptfunc} with $\partial_{\mu} V_{\mu}^3$ on the rhs vanishes because of \pref{VCC}. On the other hand, performing a partial integration on the rhs we find the relation
\begin{equation}\label{VCCrelation3pt}
\partial_{t'} C_{3,V^3_{4}}(\vec{q},t,t') = i \sum_{k=1}^3 q_k C_{3,V^3_{k}}(\vec{q},t,t')\,.
\end{equation}
This is an identity for all momentum transfer and all times $t,t'$. It  provides a nontrivial relation for the correlation functions  that will be used to test the ChPT results for these correlation functions, see section \ref{sec:ffchpt}. 
For vanishing momentum transfer \pref{VCCrelation3pt}  simplifies to
\begin{equation}\label{VCCrelation3ptZERO}
0 = \partial_{t'} C_{3,V^3_{4}}(0,t,t').
\end{equation}
Thus, the 3-pt function is independent of $t'$ and we find
\begin{equation}\label{VCCrelation3ptZERO2}
 C_{3,V^3_{4}}(0,t,t') = Q^3  C_{2}(0,t).
\end{equation}

%========================
\section{Excited-state analysis}\label{sec:excitedstates} 
%========================

\subsection{Preliminaries}

Lattice calculations of the form factors along the lines sketched in the previous section hinge on the asymptotic values of the ratios $R_{\mu}(\vec{q},t,t')$ once all time separations $t,t'$ and $t-t'$ are taken to infinity.  In practice the time separations are always finite and restricted to rather modest values well below 2 fm. Therefore, in all these cases $t'$ and $t-t'$ are smaller than 1 fm, and this is far from being asymptotically large. 

For finite time separations the 2-pt and 3-pt functions not only contain the contributions of the lowest-lying single nucleon state, but also of excited states with the same quantum numbers as the nucleon. This excited-state contribution enters the form factors too if $R_{\mu}(\vec{q},t,t')$  instead of $ \Pi_{\mu}(\vec{q})$ is used to compute the form factors. In other words we obtain {\rm effective} form factors $G^{\rm eff}_{\rm E}(Q^2,t,t')\,, {G}^{\rm eff}_{\rm M}(Q^2,t,t')$ including an excited-state contamination instead of the actual form factors we are interested in. In general we expect the effective form factors to be of the form 
\begin{eqnarray}
G^{\rm eff}_{\rm X}(Q^2,t,t')\, = \,G_{\rm X}(Q^2)\bigg[ 1 + \Delta G_{\rm X}(Q^2,t,t')\bigg],\quad X\,=\,E,M\,,\label{EffGX}
\end{eqnarray}
with the excited-state contribution $\Delta G_{\rm X}(Q^2,t,t')$ that vanish for $t,t',t-t'\rightarrow \infty$. 

For pion masses as small as in Nature one can expect two-particle $N\pi$ states to cause the dominant excited-state contamination for large but finite time separations. This expectation rests on the naive observation that the energy gaps between the $N\pi$ states and the single nucleon ground state are smaller than those one expects from true resonance states like the Roper resonance. This not only requires small pion masses but also sufficiently large spatial volumes such that the discrete spatial momenta imply small energies for the lowest-lying $N\pi$ states. Volumes with $M_{\pi}L\simeq 4$ already fulfil this criterion \cite{Bar:2017kxh}.

In this section we derive formulae that capture the $N\pi$-state contamination in the 2-pt and 3-pt functions, the ratio $R_{\mu}$ and the effective form factors. In these expressions the $N\pi$-state contamination is parameterised in terms of coefficients stemming from ratios of various matrix elements with $N\pi$ states as initial and/or final states. In the next section ChPT will be used to compute these coefficients, making the following results useful in practice. 

\subsection{$\np$ states in the 2-pt function}
The results for the 2-pt function have already been derived in Ref.\ \cite{Bar:2016uoj} because the 2-pt functions also enters the calculation of the axial form factors. For the readers convenience we briefly summarise the results here.

Performing the standard spectral decomposition in \pref{Def:2ptfunc} the 2-pt function can be written as a sum of various contributions,
\begin{eqnarray}\label{2ptDecomp}
C_2(\vec{q},t) & = & C^N_2(\vec{q},t) + C^{\np}_2(\vec{q},t)+\ldots\,.
\end{eqnarray}
The first two terms on the right hand side denote the SN and the $N\pi$ contributions. The ellipsis refers to contributions by excited states other than two-particle $N\pi$ states. We assume these to be small  and negligible compared to the ones explicitly given. 

The SN contribution is given by
\begin{equation}\label{SNcontr}
C^N_2(\vec{q},t)=\frac{1}{2E_{N,\vec{q}}}\;|\langle 0|N(0)|N(-\vec{q})\rangle|^2 e^{-E_{N,-\vec{q}}\, | t |} \,.
\end{equation}
Here $|N(-\vec{q})\rangle$ denotes the state for a moving nucleon with momentum $-\vec{q}$. The interpolating field $N(0)$ also excites $N\pi$  states with the same quantum numbers as the nucleon, thus we obtain the non-vanishing $N\pi$ contribution 
\begin{eqnarray}\label{Npicontr}
C^{\np}_2(t)&=&\frac{1}{L^3}\;\sum_{\vec{p}}\frac{1}{4E_{N,\vec{r}} E_{\pi,\vec{p}}}\,
|\langle 0|N(0)|N(\vec{r}) \pi(\vec{p})\rangle|^2 e^{-E_{\rm tot}|t|}\,.
\end{eqnarray}
The sum runs over all pion momenta that are compatible with the periodic boundary conditions.\footnote{
As usual the spatial volume is assumed to be finite with spatial extent $L$ and periodic boundary conditions are imposed for all spatial directions. The time extent is taken infinite, for simplicity.} The nucleon momentum is fixed to $\vec{r}=-\vec{q}-\vec{p}$. $E_{\rm tot}$ is the total energy of the $N\pi$ state. For weakly interacting pions $E_{\rm tot}$ equals approximately the sum $E_{N,\vec{r}}+ E_{\pi,\vec{p}}$ of the individual energies of the nucleon and the pion. 

Since the leading SN contribution is nonzero we can rewrite eq.\ \pref{2ptDecomp} as
\begin{eqnarray}
C_2(\vec{q},t) & =& C^N_2(\vec{q},t)\left\{1+ \sum_{\vec{p}} d(\vec{q},\vec{p})e^{-\Delta E(\vec{q},\vec{p}) t}\right\}\,.\label{DefC2Npcontr}
\end{eqnarray}
The coefficient $d(\vec{q},\vec{p})$ is essentially the ratio of the matrix elements appearing in eqs.\ \pref{Npicontr} and \pref{SNcontr}. The first argument in the coefficients $d(\vec{q},\vec{p})$ refers to the injected momentum $\vec{q}$, while the second one refers to the pion momentum, which we always label by $\vec{p}$. 

The energy gap $\Delta E(\vec{q},\vec{p})$ reads
\begin{equation}\label{Egap2pt}
\Delta E(\vec{q},\vec{p}) = E_{\pi,\vec{p}} + E_{N,\vec{q}+\vec{p}} - \ENq\,.
\end{equation}
The sum of the pion and nucleon energy is just the total energy of the two-particle state where at least one of the two particles carries the opposite injected momentum $-\vec{q}$. For vanishing pion momentum this is the nucleon. Alternatively $\vec{p}=-\vec{q}$, so the nucleon is at rest and the pion carries momentum $-\vec{q}$.

As mentioned before, \pref{Egap2pt} ignores the interaction energy between nucleon and pion. In the next section we compute the 2-pt function in ChPT, and to LO we will recover the result  \pref{Egap2pt} for the energy gap. Deviations due to the nucleon-pion interaction will show up at higher order in the chiral expansion.

The 2-pt function enters the generalised ratio $R_{\mu}(\vec{q},t,t')$ in \pref{DefRatio}. Introducing the short hand notation $\sqrt{\Pi C_2}$ for the square root expression in \pref{DefRatio} and expanding in powers of small quantities we obtain
\begin{equation}\label{ratio2ptGeneric}
\frac{1}{C_2(0,t)}\sqrt{\Pi C_2} = \frac{1}{C^N_2(0,t)}\sqrt{\Pi C^N_2} \left\{ 1+ \frac{1}{2} Y(\vec{q},\vec{p})\right\}\,,
\end{equation}
where the function $Y(\vec{q},\vec{p})$ contains the $\np$-state contribution,
\begin{eqnarray}
Y(\vec{q},\vec{p})&=&\sum_{\vec{p}}\Bigg(d(\vec{q},\vec{p})  \left\{ e^{-\Delta E(\vec{q},\vec{p}) (t-t')} - e^{-\Delta E(\vec{q},\vec{p}) t'} - e^{-\Delta E(\vec{q},\vec{p}) t}\right\}\nn\\
& & \phantom{\sum_{\vec{p}}} - d(0,\vec{p})  \left\{ e^{-\Delta E(\vec{0},\vec{p}) (t-t')} - e^{-\Delta E(\vec{0},\vec{p}) t'} + e^{-\Delta E(\vec{0},\vec{p}) t}\right\}\Bigg)\,.\label{DefY}
\end{eqnarray}

\subsection{$N\pi$ states in the vector current 3-pt function }

In analogy to the 2-pt function we write for the 3-pt function
\begin{eqnarray}
C_{3,\mu}(\vec{q},t,t') & = & C^N_{3,\mu}(\vec{q},t,t') + C^{\np}_{3,\mu}(\vec{q},t,t')\,.\label{DefC3NpcontrA}
\end{eqnarray}
Here $C^N_{3,\mu}$ denotes the SN result for the 3-pt function, and we assume that this contribution is nonzero. If that is the case we can alternatively write
\begin{eqnarray}
C_{3,\mu}(\vec{q},t,t') & =& C^N_{3,\mu}(\vec{q},t,t')\bigg(1+ Z_{\mu}(\vec{q},t,t')\bigg) \,,\label{DefZ}
\end{eqnarray}
with 
\begin{equation}\label{IntroZmu}
Z_{\mu}(\vec{q},t,t') =\frac{C^{N\pi}_{3,\mu}(\vec{q},t,t')}{C^N_{3,\mu}(\vec{q},t,t')}\,.
\end{equation} 
The generic form for the ratio $Z_{\mu}(\vec{q},t,t')$ is found as
\begin{eqnarray}
Z_{\mu}(\vec{q},t,t') & = &  \sum_{\vec{p}} b_{\mu}(\vec{q},\vec{p}) e^{-\Delta E(0,\vec{p}) (t-t')}+\sum_{\vec{p}}\tilde{b}_{\mu}(\vec{q},\vec{p}) e^{-\Delta E(\vec{q},\vec{p}) t'} \nn\\
& & +  \sum_{\vec{p}} c_{\mu}(\vec{q},\vec{p}) e^{-\Delta E(0,\vec{p}) (t-t')}e^{-\Delta E(\vec{q},\vec{p}) t'}\label{DefZmu}\\
& & +  \sum_{\vec{p}} \tilde{c}_{\mu}(\vec{q},\vec{p}) e^{-\Delta E(0,\vec{p}) (t-t')}e^{-\Delta E(\vec{q},\vec{p}-\vec{q}) t'}\,,\nn
\end{eqnarray}
with the energy gaps specified in eq.\ \pref{Egap2pt}. 
The coefficients $b_{\mu}(\vec{q},\vec{p}) ,\tilde{b}_{\mu}(\vec{q},\vec{p}), c_{\mu}(\vec{q},\vec{p}), \tilde{c}_{\mu}(\vec{q},\vec{p})$ in \pref{DefZmu} contain ratios of matrix elements involving the nucleon interpolating fields and the vector current. For example, the coefficient $b_{\mu}(\vec{q},\vec{p})$ contains the matrix element $\langle N\pi |V_{\mu}^3|N \rangle$ with the $N\pi$ state as the final state.  Similarly, $\tilde{b}_{\mu}(\vec{q},\vec{p})$ contains the matrix element with the $N\pi$ state as the initial state. Together the $b_{\mu}(\vec{q},\vec{p})$ and $\tilde{b}_{\mu}(\vec{q},\vec{p})$ contribution forms the excited-to-ground-state contribution. Similarly, the $c_{\mu}(\vec{q},\vec{p})$ and $\tilde{c}_{\mu}(\vec{q},\vec{p})$ contributions are called the excited-to-excited-state contribution, since it involves the matrix elements with $N\pi$ states as initial and final states. The first one captures the contribution with the nucleon absorbing the injected momentum at $t'$, while in the second one the pion absorbs it. The time dependence of these processes is slightly different, except for the special case where the momentum transfer vanishes.

As before, the sums in \pref{DefZmu} run over the momentum of the pion in the $N\pi$ state. The associated nucleon momentum is fixed by momentum conservation and the kinematic setup we have chosen.

For the calculation of the form factors according to \pref{AsympValueRePik}, \pref{AsympValueImPik} we need the expressions for the real and imaginary parts of the 3-pt function in \pref{DefC3NpcontrA}. If we consider these and rewrite them as before we obtain
\begin{eqnarray}
{\rm Re}\,C_{3,\mu}(\vec{q},t,t') & = & {\rm Re}\,C^N_{3,\mu}(\vec{q},t,t')\bigg(1+ Z^{\rm re}_{\mu}(\vec{q},t,t')\bigg) \,,\label{DefZre}\\
{\rm Im}\,C_{3,\mu}(\vec{q},t,t') & = & {\rm Im}\,C^N_{3,\mu}(\vec{q},t,t')\bigg(1+ Z^{\rm im}_{\mu}(\vec{q},t,t')\bigg) \,.\label{DefZim}
\end{eqnarray}
The $Z^{\rm re}_{\mu}$ and $Z^{\rm im}_{\mu}$ are the ratios of the real and imaginary parts of the $N\pi$ contribution and the SN contribution,
\begin{eqnarray}
Z^{\rm re}_{\mu}(\vec{q},t,t') &=& \frac{{\rm Re}\,C^{N\pi}_{3,\mu}(\vec{q},t,t')}{{\rm Re}\,C^N_{3,\mu}(\vec{q},t,t')}\,,\\
Z^{\rm im}_{\mu}(\vec{q},t,t') &=& \frac{{\rm Im}\,C^{N\pi}_{3,\mu}(\vec{q},t,t')}{{\rm Im}\,C^N_{3,\mu}(\vec{q},t,t')}\,.
\end{eqnarray}
Note that these are not the real and imaginary parts of $Z_{\mu}$. The general structure of $Z^{\rm re}_{\mu}, Z^{\rm im}_{\mu}$  reads ($x=$ re or im)
\begin{eqnarray}
Z^{\rm x}_{\mu}(\vec{q},t,t') & = &   \sum_{\vec{p}} b^x_{\mu}(\vec{q},\vec{p}) e^{-\Delta E(0,\vec{p}) (t-t')}+\tilde{b}^x_{\mu}(\vec{q},\vec{p}) e^{-\Delta E(\vec{q},\vec{p}) t'} \nn\\
& & +  \sum_{\vec{p}} c^x_{\mu}(\vec{q},\vec{p}) e^{-\Delta E(0,\vec{p}) (t-t')}e^{-\Delta E(\vec{q},\vec{p}) t'}\,,\nn\\
& & +  \sum_{\vec{p}} \tilde{c}^x_{\mu}(\vec{q},\vec{p}) e^{-\Delta E(0,\vec{p}) (t-t')}e^{-\Delta E(\vec{q},\vec{p}-\vec{q}) t'}.\label{DefC3Npcontr}
\end{eqnarray}
Here too one should keep in mind that $b_{\mu}\neq b^{\rm re}_{\mu}+ib^{\rm im}_{\mu}$ and analogously for the other coefficients.

It is worth pointing out a crucial difference to the analogous expressions for the axial form factors. Comparing \pref{DefZmu} with eq.\ (3.11) in \cite{Bar:2018xyi} we observe that the $N\pi$ state contributions $b_{\mu},\tilde{b}_{\mu}, c_{\mu}$ appear in both cases. The remaining $\tilde{c}_{\mu}$ contribution, however, is absent. Instead,  additional ones, parameterised by coefficients $a_{\mu}, \tilde{a}_{\mu}$, appear in case of the axial vector current. The reason for this lies in the different symmetry properties of the vector and axial vector currents and is easily understood. The axial vector current directly couples to a pion, thus at operator insertion time $t'$ it can directly create a pion that travels to and gets destroyed at the sink ($a_{\mu}$ contribution). Alternatively the axial vector can directly destroy a pion created at the source  ($\tilde{a}_{\mu}$ contribution). The direct pion coupling to the vector current must involve at least two pions, thus these two contributions are absent in \pref{DefZmu}. However, the vector current can destroy a pion stemming from the source and at the same time create a pion that subsequently travels to the sink. This is exactly the $\tilde{c}_{\mu}$ contribution in \pref{DefZmu}.\footnote{Jumping ahead, the difference is obvious in the Feynman diagrams contributing to the coefficients, see fig.\ \ref{fig:Npidiags3pt}, diagram m) to p) compared with diagrams (m) and (n) in fig.\ 3 of Ref.\ \cite{Bar:2018xyi}. Note that the latter ones are tree diagrams. Hence, their contribution to the $N\pi$ contamination in the axial form factors is found to be much larger than the other contributions coming from one-loop diagrams. Since this contribution is missing for the vector current we expect a smaller $N\pi$ contamination for the electromagnetic form factors. This expectation is confirmed in section \ref{sect:impact}.} The direct analogue of the $a_{\mu}, \tilde{a}_{\mu}$ contribution stems from two pions propagating between either source or sink and the operator. This, however, is a 3-particle $N\pi\pi$ contribution and beyond the scope of this paper.

\subsection{The ratios and effective form factors}

Forming the ratio of the 3-pt function with the 2-pt function we obtain the total result for the ratios,
\begin{eqnarray}
R_{\mu}(\vec{q},t,t')&=& \Pi_{\mu}(\vec{q}) \Bigg(1+ Z_{\mu}(\vec{q},t,t') + \frac{1}{2} Y(\vec{q},t,t')\Bigg)\,,\label{NpiConttot}
\end{eqnarray}
with $\Pi_{\mu}(\vec{q})$ referring to the asymptotic values of the ratios given in \pref{Pimu}. Obviously the ratios approach the correct asymptotic values, by construction. 

Eqs.\ \pref{AsympValuePi4} - \pref{AsympValueImPik} require to take the real or imaginary part of \pref{NpiConttot}. We will later find that, to the order we are working here, the result for the 2-pt function is real. In that case the real and imaginary parts of \pref{NpiConttot} are given by
\begin{eqnarray}
{\rm Re}\, R_{\mu}(\vec{q},t,t')&=& {\rm Re}\, \Pi_{\mu}(\vec{q}) \Bigg(1+ Z^{\rm re}_{\mu}(\vec{q},t,t') + \frac{1}{2} Y(\vec{q},t,t')\Bigg)\,,\label{NpiConttotRE}\\
{\rm Im}\, R_{\mu}(\vec{q},t,t')&=& {\rm Im}\, \Pi_{\mu}(\vec{q}) \Bigg(1+ Z^{\rm im}_{\mu}(\vec{q},t,t') + \frac{1}{2} Y(\vec{q},t,t')\Bigg)\,.\label{NpiConttotIM}
\end{eqnarray}
As discussed before, the form factors are obtained from the asymptotic values by multiplication with trivial kinematic factors, see eqs.\ \pref{AsympValuePi4} - \pref{AsympValueImPik}. If the ratios at finite times $t,t'$ are used we obtain {\em effective form factors} that contain the $N\pi$ excited-state contribution. Explicitly,
\begin{eqnarray}
\GEf^{\rm eff}(Q^2,t,t') &=& \GE(Q^2) \Bigg(1+ Z^{\rm re}_{4}(\vec{q},t,t') + \frac{1}{2} Y(\vec{q},t,t')\Bigg)\,,\label{GEeffV4}\\
\GM^{\rm eff}(Q^2,t,t') &=& \GM(Q^2)\Bigg(1+ Z^{\rm re}_{i}(\vec{q},t,t') + \frac{1}{2} Y(\vec{q},t,t')\Bigg)\,,\label{GMeffVi}\\
\GEi^{\rm eff}(Q^2,t,t') &=& \GE(Q^2)\Bigg(1+ Z^{\rm im}_{i}(\vec{q},t,t') + \frac{1}{2} Y(\vec{q},t,t')\Bigg)\,.\label{GEeffVi}
\end{eqnarray}
The additional subscript for the electric form factors indicates what formula has been used to obtain the effective form factor, eq.\ \pref{AsympValuePi4} or eq.\ \pref{AsympValueImPik}. 
From these the familiar estimators for the form factors, the {\em midpoint} or the {\em plateau} estimates are defined in the usual way. For example, the former ones are given by
\begin{eqnarray}
G_{\rm X}^{\rm mid}(Q^2,t) & = & G_{\rm X}^{\rm eff}(Q^2,t,t'=t/2) \,.
\end{eqnarray}
For the plateau estimates the operator insertion time $t'$ assumes the value such that the effective form factor is extremal. For small momentum transfers one finds $t'\approx t/2$. Thus, for simplicity, we only consider the midpoint estimates in the following.

Note that the Ward identity \pref{VCCrelation3ptZERO2} implies 
\begin{eqnarray}
\GEf^{\rm eff}(Q^2=0,t,t') &=& \GE(Q^2=0) \,=\,1\,.
\end{eqnarray}
Thus, vector current conservation automatically results in the correct result for the electric form factor for vanishing momentum transfer, irrespective of the excited-state contribution to the correlation functions.

%========================
\section{$N\pi$-state contribution in ChPT}\label{sec:ffchpt} 
%========================

\subsection{General remarks}

For large times $t,t',$ pion physics dominates the correlation functions that we defined in the previous section. In that case ChPT, the low-energy effective theory of QCD \cite{Weinberg:1978kz,Gasser:1983yg,Gasser:1984gg}, is expected to provide good estimates for them. In particular, forming the ratio $R_{\mu}$ we obtain ChPT results for the various  coefficients parameterising the $N\pi$ contamination in the effective form factors. 

Such ChPT calculations have been performed for a variety of nucleon observables, for example the nucleon mass, nucleon charges and moments of parton distribution functions, see \cite{Bar:2015zwa,Tiburzi:2015tta,Bar:2016uoj,Bar:2016jof}. Ref.\ \cite{Bar:2018xyi} reports an analogous ChPT calculation for the $N\pi$ contamination in the effective axial form factors of the nucleon. The computation presented here is completely analogous, and differs only in some details stemming from the different expressions for the vector and axial vector currents.

The calculation is done in covariant ChPT \cite{Gasser:1987rb,Becher:1999he} to LO. The ChPT setup with the Feynman rules and the chiral expressions for the vector current and the nucleon interpolating fields are summarised in appendix \ref{appFeynmanRules}. For some more details the reader is referred to the reviews \cite{Bar:2017kxh,Bar:2017gqh}.

To the order we are working here the results for the various coefficients depend on three low-energy coefficients (LECs) only: the chiral limit values of the pion decay constant $f$,  the axial charge $g_A$ and the difference $\mu_{p-n}\equiv \mu_p-\mu_n$ of the proton and neutron's magnetic moments. Since all these are known phenomenologically very well the LO ChPT results are very predictive.  

\subsection{The $N\pi$-state contribution}\label{ssect:Npicontribution}

%=====================
\begin{figure}[t]
\begin{center}
\includegraphics[scale=0.5]{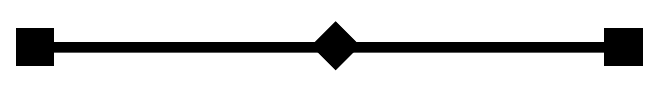}\\
\caption{Feynman diagram for the leading single nucleon contribution in the  vector current 3-pt function. Squares represent the nucleon interpolating fields at times $t$ and $0$, and the diamond stands for the vector current at insertion time $t'$.
The solid lines represent a nucleon propagator, and a momentum $-\vec{q}$ is injected at $t'$.
}
\label{fig:diagsSN}
\end{center}
\end{figure}
%=====================

\begin{figure}[t]
%\begin{center}
\includegraphics[scale=0.45]{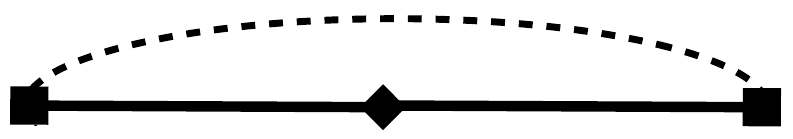}\hspace{0.3cm}\includegraphics[scale=0.45]{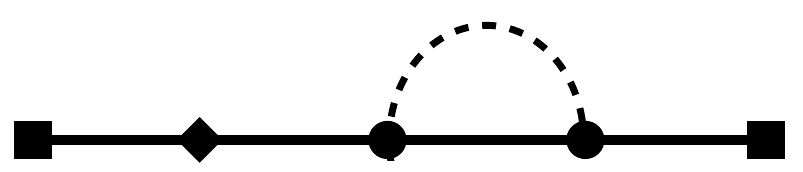}\hspace{0.3cm}\includegraphics[scale=0.45]{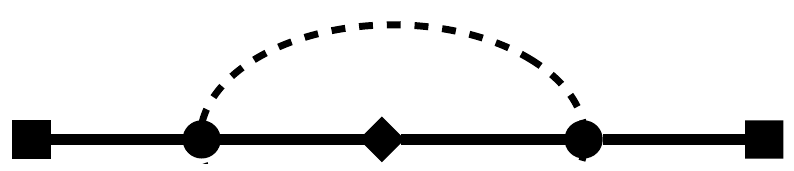}\hspace{0.3cm}\includegraphics[scale=0.45]{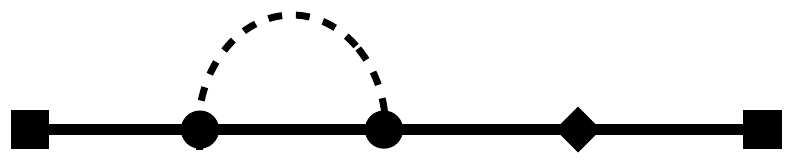}\\
a)\hspace{3.5cm} b)\hspace{3.5cm} c)\hspace{3.5cm} d)\\[3ex]
\includegraphics[scale=0.45]{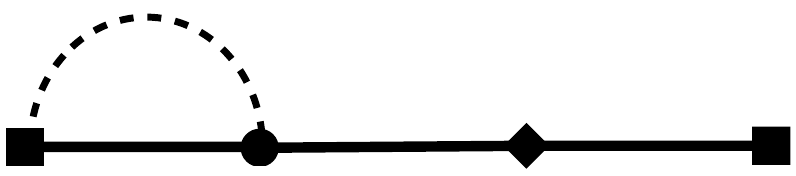}\hspace{0.3cm}\includegraphics[scale=0.45]{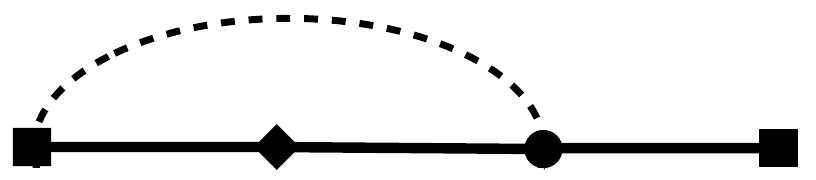}\hspace{0.3cm}\includegraphics[scale=0.45]{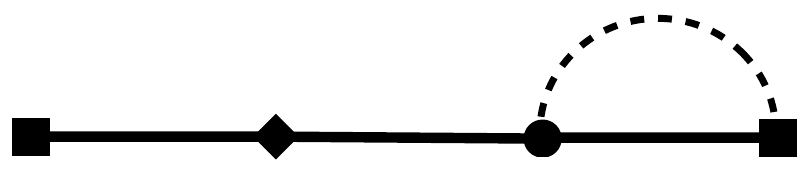}\hspace{0.3cm}\includegraphics[scale=0.45]{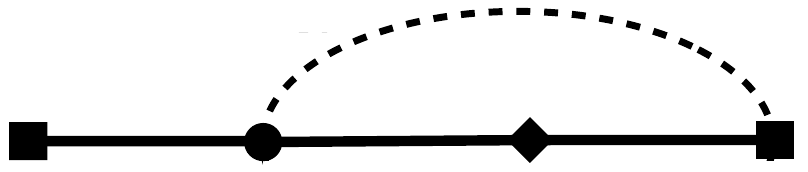}\\
e)\hspace{3.5cm} f)\hspace{3.5cm} g)\hspace{3.5cm} h) \\[3ex]
\includegraphics[scale=0.45]{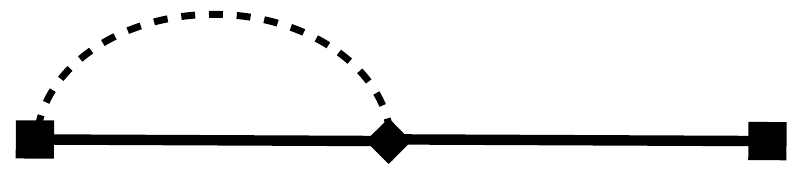}\hspace{0.3cm}\includegraphics[scale=0.45]{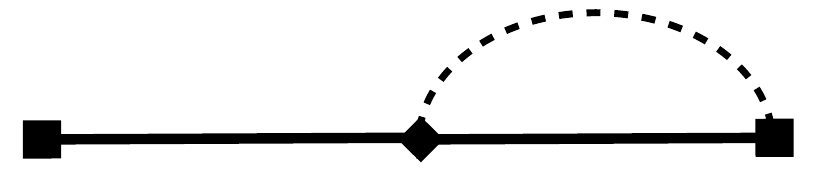}\hspace{0.3cm}\includegraphics[scale=0.45]{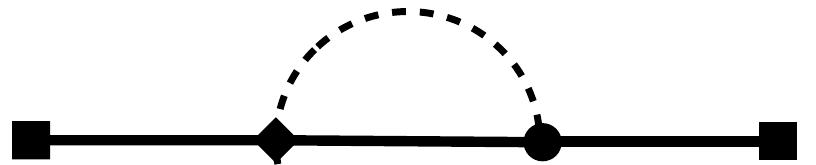}\hspace{0.3cm}\includegraphics[scale=0.45]{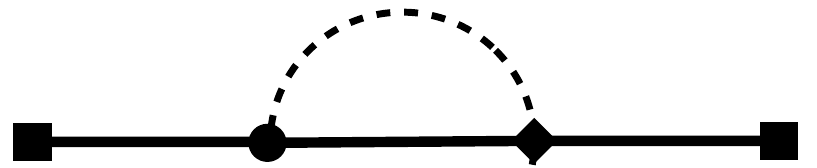}\\
i)\hspace{3.5cm} j)\hspace{3.5cm} k)\hspace{3.5cm} l)\\[3ex]
\includegraphics[scale=0.45]{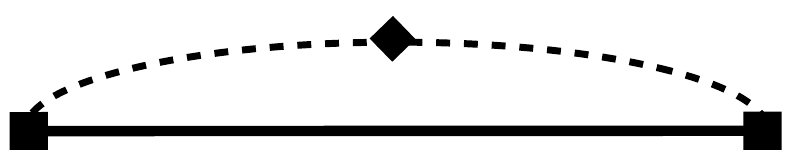}\hspace{0.3cm}\includegraphics[scale=0.45]{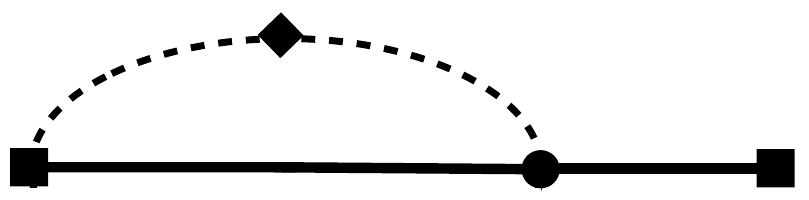}\hspace{0.3cm}\includegraphics[scale=0.45]{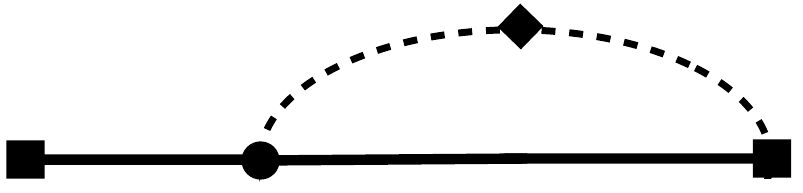}\hspace{0.3cm}\includegraphics[scale=0.45]{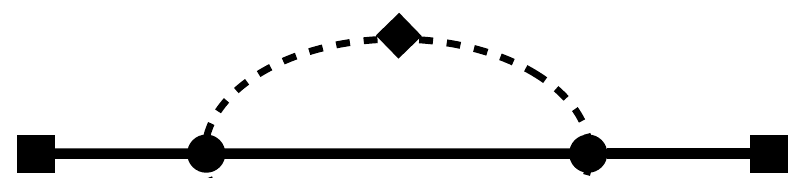}\\
m)\hspace{3.5cm} n)\hspace{3.5cm} o)\hspace{3.5cm} p)\\[0.3ex]
\caption{Feynman diagrams for the leading $N\pi$ contribution in the vector current 3-pt function. Circles represent a vertex insertion at an intermediate space-time point, and an integration over this point is implicitly assumed. The dashed lines represent pion propagators. }
\label{fig:Npidiags3pt}
%\end{center}
\end{figure}

With the Feynman rules given in appendix \ref{appFeynmanRules} it is straightforward to draw the leading diagrams for the correlation functions of interest. Figure \ref{fig:diagsSN} shows the single diagram for the leading SN contribution in the vector current 3-pt function. The leading $N\pi$ contribution stems from the sixteen loop diagrams depicted in figure \ref{fig:Npidiags3pt}. The calculation of these diagrams is a standard task in ChPT. Five more diagrams are needed for the 2-pt function, but the results can be taken from Ref.\ \cite{Bar:2018xyi}. 
With the expressions for the 2-pt and 3-pt functions we form the ratio $R_{\mu}$ and read off the coefficients $b^x_{\mu}(\vec{q},\vec{p}) ,\tilde{b}^x_{\mu}(\vec{q},\vec{p}), c^x_{\mu}(\vec{q},\vec{p}), \tilde{c}^x_{\mu}(\vec{q},\vec{p})$.

Following \cite{Bar:2018xyi} it is useful to separate the coefficients into a universal part and a ,,reduced" coefficient, for instance
\begin{equation}\label{DefRedCoeff}
d(\vec{q},\vec{p}) =   \frac{1}{8 (fL)^2 \Epip L} D(\vec{q},\vec{p})
\end{equation}
and analogously for the other coefficients $b^x_{\mu}(\vec{q},\vec{p}) ,\tilde{b}^x_{\mu}(\vec{q},\vec{p}), c^x_{\mu}(\vec{q},\vec{p}), \tilde{c}^x_{\mu}(\vec{q},\vec{p})$.
The universal factor collects the spatial volume $L^3$ in the dimensionless combinations $\Epip L$ and $fL$. This factor is expected to appear and stems from the loop diagrams in figure \ref{fig:Npidiags3pt}.

The reduced coefficients, denoted by capital letters, are dimensionless functions involving the nucleon and pion energies and momenta and the injected momentum transfer.
The expressions are rather cumbersome in full covariant form. They simplify significantly if we perform the non-relativistic (NR) expansion for the nucleon energy, 
\begin{equation}\label{NRexpansion}
\ENq = M_N+\frac{\vec{q}^{\,2}}{2M_N} ,
\end{equation}
and keep only the first two terms. For practical uses this is sufficient. 
For example, the expansion of the reduced coefficient $D(\vec{q},\vec{p})$, defined in \pref{DefRedCoeff}, reads
\begin{equation}\label{DefNRExp}
D(\vec{q},\vec{p}) =D^{\infty}(\vec{q},\vec{p}) + \frac{\Epip}{M_N}D^{\rm corr}(\vec{q},\vec{p})\,.
\end{equation}
$D^{\infty}(\vec{q},\vec{p}) $ gives the value if the nucleon mass were infinite, $D^{\rm corr}(\vec{q},\vec{p})$ the O($1/M_N$) correction. Both were calculated in Ref.\ \cite{Bar:2018xyi} with the following results
\begin{eqnarray}
D^{\infty}(\vec{q},\vec{p}) & = & 3g_A^2\frac{ p^2}{\Epip^2}\,,\\
D^{\rm corr}(\vec{q},\vec{p}) & = & 3g_A\frac{g_A M_{\pi}^2 (p^2+2pq)-\Epip^2 (p^2+pq)}{\Epip^4}\,,
\end{eqnarray}
where we used the abbreviations
\begin{eqnarray}
p^2=\vec{p}^{\,2}\,,\qquad pq=\vec{p}\cdot\vec{q}\,.
\end{eqnarray}

% Figure
%
% End figure

The main new results of this paper are the coefficients stemming from the vector current 3pt function. For the index $\mu=4$ our results for the leading NR limit coefficients read
\begin{eqnarray}
B_4^{re,\infty}(\vec{q},\vec{p}) & =& 4 g_A^2 \left(\frac{p^2}{\Epip^2} - \frac{p^2-pq}{\Epis^2}  \right)\,,\\
\tilde{B}_4^{re,\infty}(\vec{q},\vec{p}) & =& 4 g_A^2 \left(\frac{p^2}{\Epip^2} - \frac{p^2+pq}{\Epir^2}  \right)\,,\\
C_4^{re,\infty}(\vec{q},\vec{p}) & =& - g_A^2 \frac{p^2}{\Epip^2}\,,\\
\tilde{C}_4^{re,\infty}(\vec{q},\vec{p}) & =&2 g_A^2 \frac{(\Epip+\Epis)(p^2-pq)}{\Epip\Epis^2}\,.\label{tildeCre4NRLIM}
\end{eqnarray}
Two different pion energies appear in these results, in particular the energies of a pion carrying the sum and the difference of $\vec{p}$ and $\vec{q}$,
\begin{eqnarray}
\vec{r} = \vec{p} + \vec{q},\qquad \vec{s} = \vec{p} - \vec{q}\,.
\end{eqnarray}

The NR expansion is slightly different for the spatial components $\mu = k$, $k=1,2,3$. The reason is that the SN contribution to the 3-pt function is O($1/M_N$), thus, it vanishes in the infinite nucleon mass limit. The $N\pi$ contribution, on the other hand, is O($1$). The coefficients we are interested in are the ratios of these two contributions. Therefore, for $\mu=k$, the  inverse power $1/M_N$ in the SN contribution shifts the NR expansion of the ratio such that powers linear in the nucleon mass appear. Explicitly, we need to define
\begin{eqnarray}
B_k^{x}(\vec{q},\vec{p})  =  \frac{M_N}{\Epip} B_k^{x,\infty}(\vec{q},\vec{p}) + B_k^{x,{\rm corr}}(\vec{q},\vec{p})\label{EnhancedNRExp}
\end{eqnarray}
and analogously for $\tilde{B}_i^{x}(\vec{q},\vec{p})$ and $\tilde{C}_i^{x}(\vec{q},\vec{p})$. Keep in mind that these coefficients diverge in the infinite nucleon mass limit, because the single nucleon contribution vanishes in this limit while the $N\pi$ contribution tends to a non-vanishing constant. 

Because of the factor $M_N/\Epip$ in \pref{EnhancedNRExp} we call these coefficients {\em O($M_N)$ enhanced}.\footnote{This kind of enhancement was already observed in the $N\pi$ contamination in the axial form factors \cite{Bar:2018xyi}.} The remaining coefficient $C^x_k$, on the other hand, starts as usual and is expanded as in \pref{DefNRExp},

For the $N\pi$ contamination in the effective magnetic form factor in \pref{GMeffVi} we need the real parts for $\mu=k=1$ or 2. For $k=1$ we find the leading coefficients
\begin{eqnarray}
B_1^{re,\infty}(\vec{q},\vec{p}) & =&+ \frac{8 g_A^2}{\mu_{p-n}} \frac{(2p_1-q_1) (p_2q_1-p_1 q_2)+\Epis^2\, p_2}{\Epis^2\, q_2}\,,\label{BkReInf}\\
\tilde{B}_1^{re,\infty}(\vec{q},\vec{p}) & =&+ \frac{8 g_A^2}{\mu_{p-n}} \frac{(2 p_1 + q_1) (p_2 q_1 - p_1 q_2)-\Epir^2\, p_2}{\Epir^2 \,q_2}\,,\label{BtkReInf}\\
\tilde{C}_1^{re,\infty}(\vec{q},\vec{p}) & =&-\frac{4 g_A^2}{\mu_{p-n}} \frac{(2 p_1 - q_1) (p_2 q_1 - p_1 q_2)}{\Epis^2 \,q_2}\,,\\
{C}_1^{re,\infty}(\vec{q},\vec{p}) &=& +\frac{g_A^2}{\mu_{p-n}} \frac{{p}^{2} q_2 + 2p_3( p_2 q_3-p_3 q_2)}{\Epip^2 \, q_2}\,.
\end{eqnarray}
Recall the short hand notation  $\mu_{p-n}=\mu_p-\mu_n$ for the difference between the magnetic moments of proton and neutron.  The corresponding results for $k=2$ are obtained by the simple substitution $q_2\rightarrow q_1, \, p_2 \rightarrow p_1$.

For the $N\pi$ contamination in the effective electric form factor in \pref{GEeffVi} we need the imaginary parts for $\mu=k$ and find 
\begin{eqnarray}
B_k^{im,\infty}(\vec{q},\vec{p}) & =&+ 8 g_A^2 \frac{({p}^{2}-{p}{q}) (2p_k - q_k)-\Epis^2 p_k}{\Epis^2\, q_k}\,,\label{BkimInf}\\
\tilde{B}_k^{im,\infty}(\vec{q},\vec{p}) & =& +8 g_A^2  \frac{({p}^{2}+{p}{q})(2p_k+q_k)-\Epir^2 p_k}{\Epir^2 \,q_k}\,,\label{BtkimInf}\\
\tilde{C}_k^{im,\infty}(\vec{q},\vec{p}) & =&- 4 g_A^2  \frac{({p}^{2}-{p}{q})(2p_k-q_k)}{\Epis^2 \,q_k}\,,\label{CtkimInf}\\
C_k^{im,\infty}(\vec{q},\vec{p}) & =&  - g_A^2  \frac{{p}^2(2 p_k + q_k)}{\Epip^2 \,q_k}\,.\label{CkimInf}
\end{eqnarray}
These results hold for $k=1,2,3$. Note that here the LEC $\mu_{p-n}$ does not appear. To the order in the NR expansion we are working to this LEC enters the effective magnetic form factor only. In the effective electric form factors it enters at O($1/M_N^2)$ and has been dropped (see appendix \ref{app:SNresults}).

The results for the correction coefficients $B_{k}^{\rm corr}(\vec{q},\vec{p}),\tilde{B}_{k}^{\rm corr}(\vec{q},\vec{p}),C_{k}^{\rm corr}(\vec{q},\vec{p})$ and $\tilde{C}_{k}^{\rm corr}(\vec{q},\vec{p})$ are cumbersome. Since the detailed expressions reveal no additional qualitative insight they are listed in appendix \ref{app:corrcoeff}. 

%
%==============================
\section{Impact on lattice calculations}\label{sect:impact}
%==============================
%

\subsection{Preliminaries}

To LO in ChPT the $N\pi$ contribution to the ratio $R_{\mu}$ and the effective form factors depends on a few LECs only, and their values can be obtained rather precisely from experimental data. Assuming these values in the ChPT results of the previous section we obtain estimates for the impact of the $N\pi$ contribution on lattice calculations of the form factors. 
The rationale for this application is the same as for the axial and pseudoscalar nucleon form factors  \cite{Bar:2018xyi,Bar:2019gfx}. The reader is referred to these references for more details. Here we merely summarise the values for the various input parameters that need to be fixed for the analysis.   

Three LECs are the chiral limit values of the pion decay constant, the axial charge and the difference of the magnetic moments of the proton and neutron.  To LO it is consistent to use the experimental values for these LECs and we set them to $f=f_{\pi}= 93$ MeV, $g_A=1.27$ and $\mu_{p-n} =4.706$ \cite{Tanabashi:2018oca}. We ignore the errors in these values since they are too small to be significant for the LO results in this paper 

Two more LECs are associated with the pion and nucleon mass. We are mainly interested in the $N\pi$ contribution in physical point simulations, so we fix the pion and nucleon masses to their (approximate) physical values $M_{\pi}=140$ MeV and $M_N=940$ MeV. 

The finite spatial volume determines the accessible spatial momenta. In practice, it is fixed by the lattice spacing and the number of lattice points in the spatial directions. Typical values in recent lattice calculations cover a range $M_{\pi}L \sim 4$ to 6, and we will assume such values in the following analysis.\footnote{A recent simulation of the PACS collaboration \cite{Shintani:2018ozy} was carried out at a larger volume with $M_{\pi}L \approx 7.4$.}  Imposing periodic boundary conditions the spatial momentum transfer can assume the values
$\vec{q}_n=(2\pi/L)\vec{n}_q$
with the vector $\vec{n}_q$ having integer valued components. 

ChPT is an expansion in the small pion mass and in small pion momenta. Therefore, we select an upper bound on the pion momentum in the $N\pi$ state. Following Refs.\ \cite{Bar:2016uoj,Bar:2016jof} we choose $|\vec{p}_n|\lesssim p_{\rm max}$ with $p_{\rm max}/\Lambda_{\chi}= 0.45$, where the chiral scale $\Lambda_{\chi}$ is equal to $4\pi f_{\pi}$. $N\pi$ states with pions satisfying this bound are called {\em low-momentum $N\pi$ states} in the following. For these we expect the LO ChPT results to work reasonably well. States with pion momenta larger than this bound are called {\em high-momentum $N\pi$ states}. These too contribute to the excited-state contamination. However, choosing all Euclidean time separations sufficiently large the contribution of the high-momentum $N\pi$ states can be made small and negligible. The results in Refs.\ \cite{Bar:2016uoj,Bar:2016jof} suggest that a separation of at least 1 fm between the operator and both source and sink is necessary for a sufficient suppression. This corresponds to source-sink separations of 2 fm or larger in the 3-pt function. Therefore, we take $t=2$ fm as our generic source-sink separation in the following.

Note that an upper bound $|\vec{p}_n|\lesssim p_{\rm max}$ translates into a number $n_{p {\rm max}}$ that depends on the spatial volume, i.e.\ on $M_{\pi}L$. The larger the volume the more discrete momenta satisfy the bound. 
A list of $n_{p, {\rm max}}$ for various volumes is given in Ref.\ \cite{Bar:2018xyi}, table 1.

%===================
\subsection{Impact on the electromagnetic form factors}\label{ssect:impactonFFs}
%===================

\begin{figure}[p]
\begin{center}
$\epsilon^{\rm eff}_{\rm E,4}(Q^2,t=2\,{\rm fm},t')$\\
\includegraphics[scale=0.65]{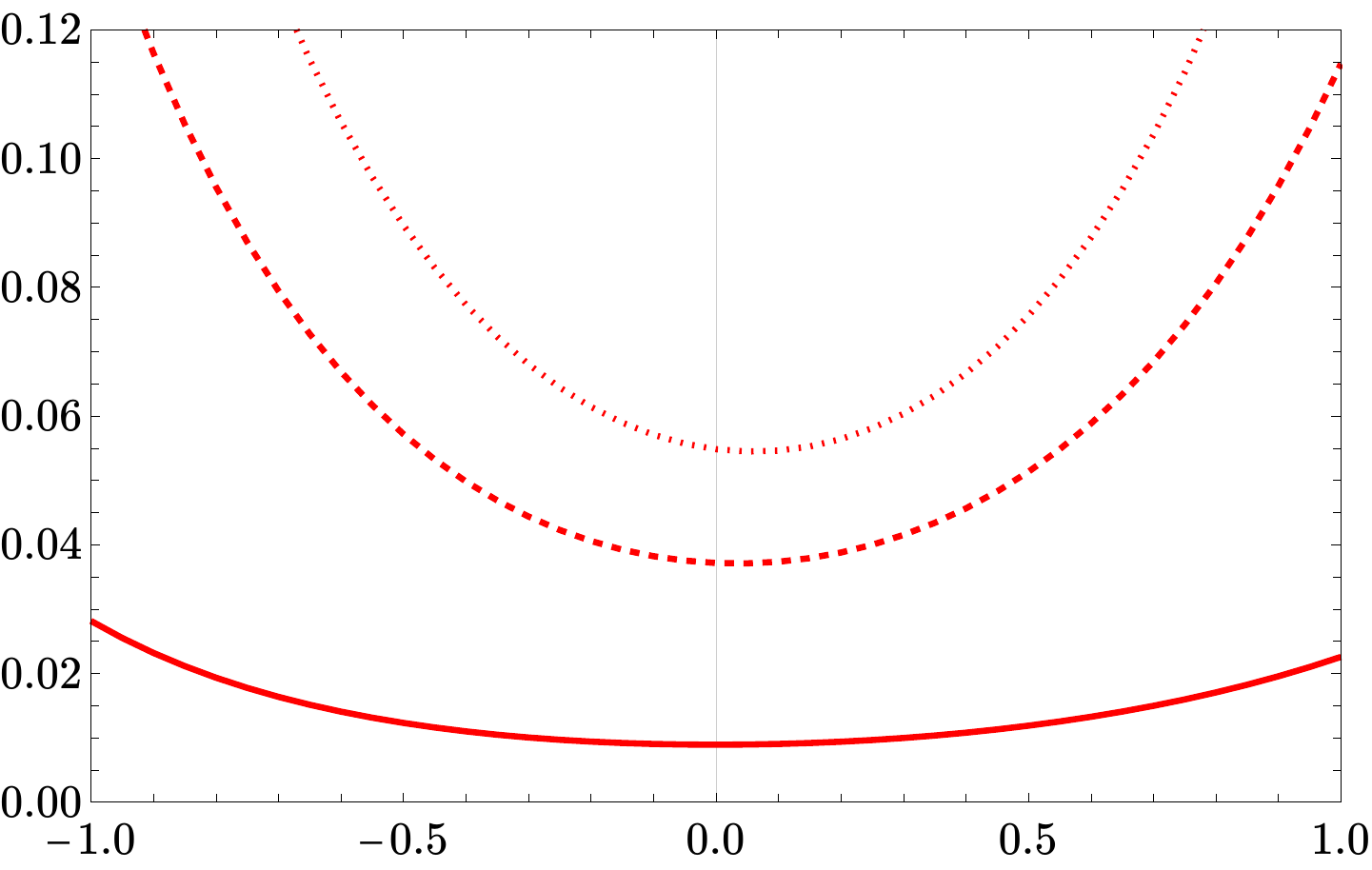}\\[4ex]
$\epsilon^{\rm eff}_{\rm M}(Q^2,t=2\,{\rm fm},t')$\\
\includegraphics[scale=0.65]{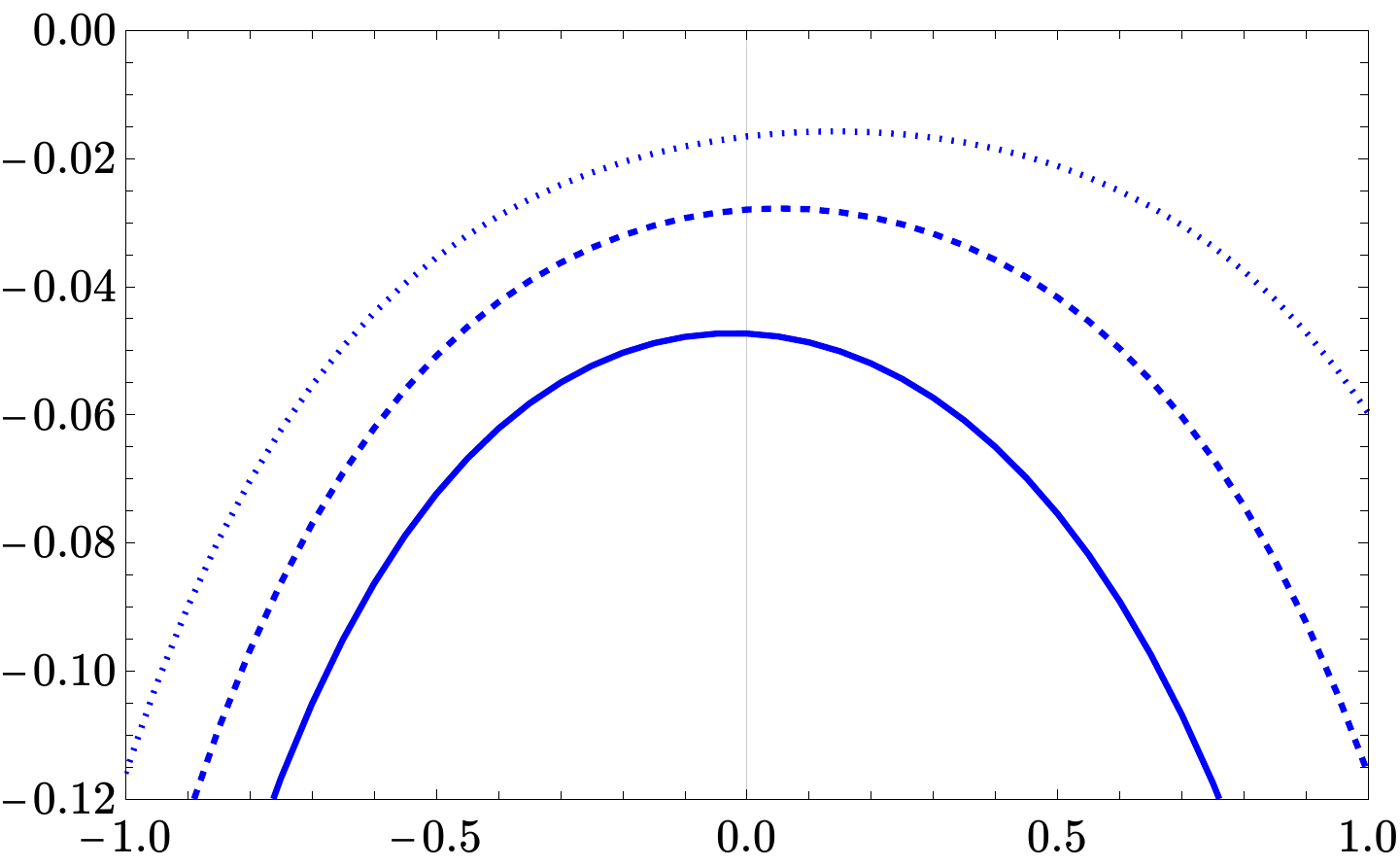}\\[4ex]
$\epsilon^{\rm eff}_{\rm E,2}(Q^2,t=2\,{\rm fm},t')$\\
\includegraphics[scale=0.65]{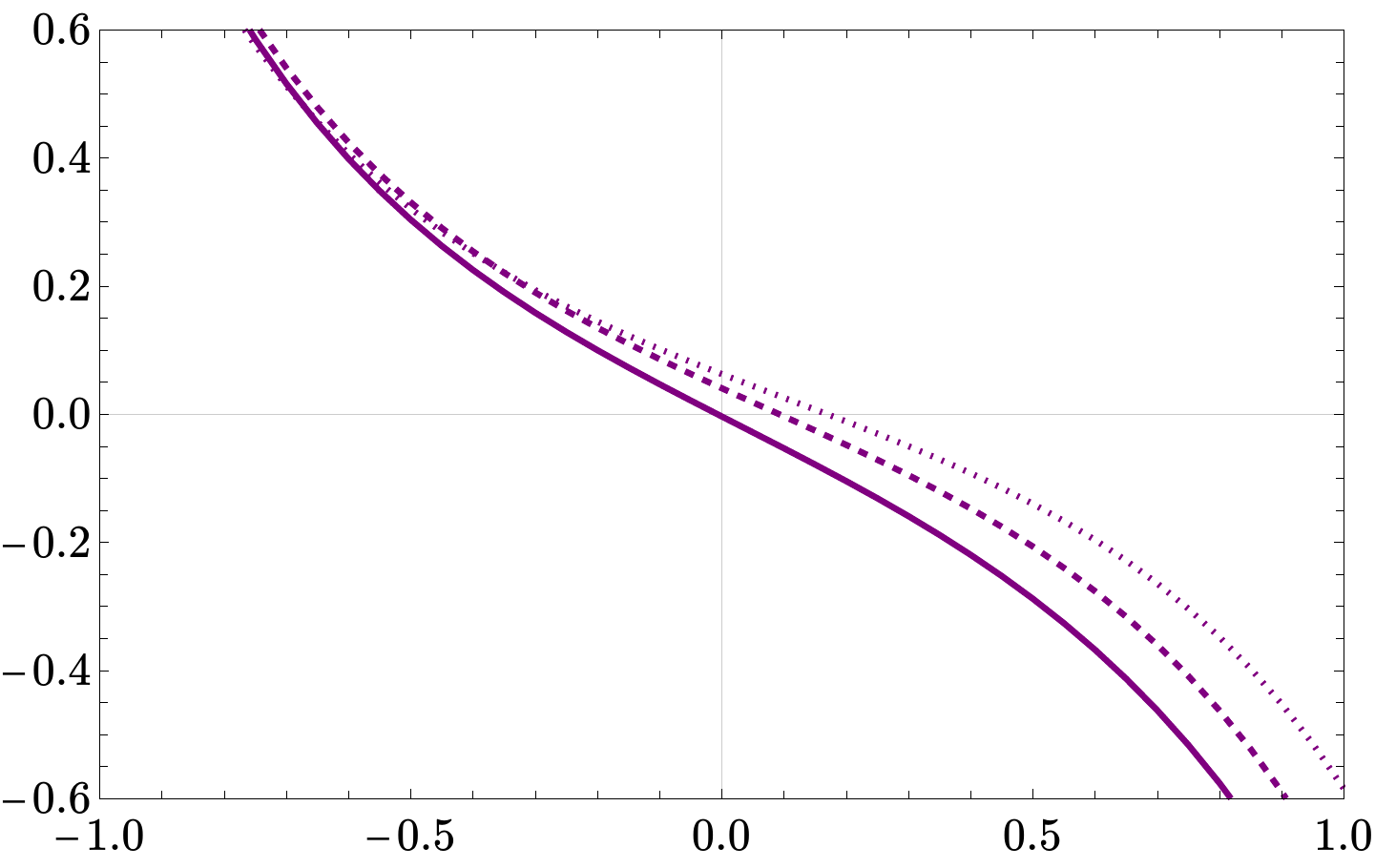}\\[1ex]
$\tau/{\rm fm}$\\[2ex]
\caption{The relative deviation $\epsilon^{\rm eff}_{\rm X}(Q^2,t=2\,{\rm fm},t')$ as a function of the shifted operator insertion time $\tau\equiv t'-t/2$ for $X=E,4$ (top panel), $X=M$ (middle) and $X=E,2$ (bottom), for three different non-zero momentum transfers with $n_q=1$ (solid lines), $n_q=6$ (dashed lines) and $n_q=12$ (dotted lines) for $M_\pi L=6$. }
\label{fig:epsEff}
\end{center}
\end{figure}

The effective form factors $G^{\rm eff}_{\rm X}(Q^2,t,t')$ in \pref{GEeffV4} -- \pref{GEeffVi}  depend on the source-sink separation $t$ and the operator insertion time $t'$. For fixed $t$ the $N\pi$ contamination is expected to be minimal for $t'\approx t/2$, at least for small momentum transfers. As a measure for the $N\pi$-state contribution we introduce the relative deviation from the true form factors,
\begin{eqnarray}
\epsilon^{\rm eff}_{\rm X}(Q^2,t,t')\equiv \frac{G^{\rm eff}_{\rm X}(Q^2,t,t')}{G_{\rm X}(Q^2)} -1\label{DefEpsilons}\,.
\end{eqnarray}
Fig.\ \ref{fig:epsEff} shows $\epsilon^{\rm eff}_{\rm X}(Q^2,t,t')$ as a function of the shifted operator insertion time $t'-t/2$ for fixed source sink separation $t=2$ fm, $M_{\pi} L=6$ and the three momentum transfers with $n_q=1$ (solid lines), $n_q=6$ (dashed lines) and $n_q=12$ (dotted lines). These values correspond to $Q_{n_q}^2\approx 0.02$, $0.13$ and $0.24$  ${\rm GeV}^2$, respectively. 

For the electric form factor obtained with the timelike component $V_4$ (top panel) the effective form factor overestimates $G_{\rm E}$, and the overestimation increases for increasing momentum transfer. For the magnetic form factor (middle panel) we observe an underestimation, which is larger for the smaller momentum transfers. In both cases we observe a cosh-like behaviour with the minimal resp.\ maximal value close to $t'=t/2$. This means that both midpoint and plateau estimate are essentially the same, as expected.

The results for the electric form factor obtained with the spatial component $V_2$ (bottom panel) is qualitatively different. Instead of a cosh-like behaviour we find an approximate sinh-like behaviour. Hence, the effective form factor does not have a plateau estimate. In addition, we observe that $\epsilon^{\rm eff}_{\rm E, 2}(Q^2,t,t')$  is significantly larger than the other two (note the different scale on the bottom panel). In other words, the $N\pi$ contamination in $G^{\rm eff}_{\rm E,2}$ is much larger than in $G^{\rm eff}_{\rm E,4}$, suggesting a preference for the latter to compute the electric form factor.

% Figure
%
% End figure

The reason for the larger $N\pi$ contamination is the O($M_N$) enhancement \pref{EnhancedNRExp} in the coefficients for the spatial components. Note  that the same enhancement is at work in the real parts that give the effective magnetic form factor. There, however, it is largely compensated by the factor $1/\mu_{p-n}$ in \pref{BkReInf}, \pref{BtkReInf}. This factor is roughly $0.2$, so the $N\pi$ contamination in $G^{\rm eff}_{\rm E,2}$ is about five time larger than in $G^{\rm eff}_{\rm M}$, in qualitative agreement with 
what we observe when comparing the middle and bottom panel in fig.\ \ref{fig:epsEff}.

\begin{figure}[t]
%\begin{center}
$\epsilon^{\rm mid}_{\rm X}(Q^2,t=2\,{\rm fm})$\\
\includegraphics[scale=0.73]{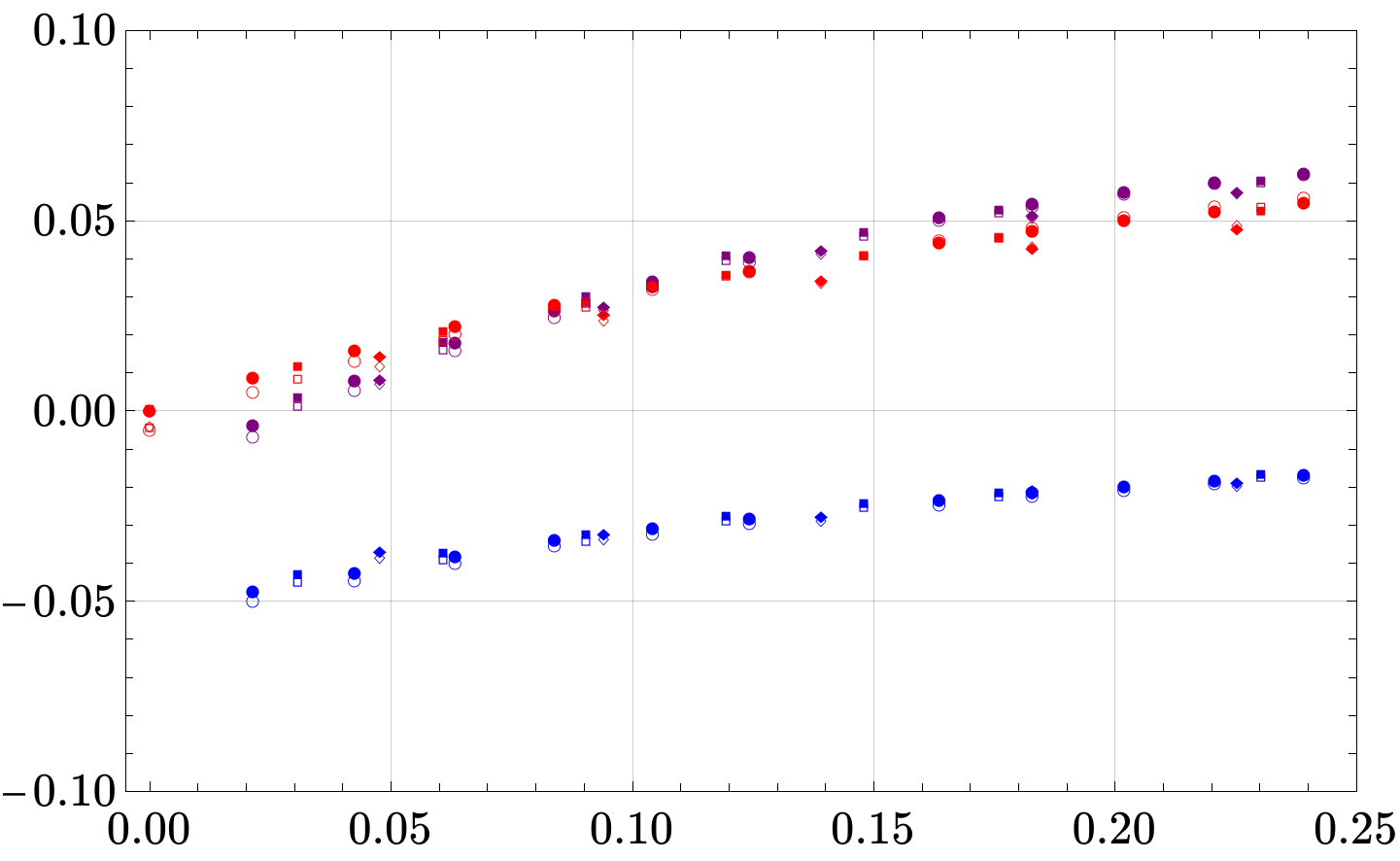}\\[0.3ex]
$Q^2/({\rm GeV})^2$\\[2ex]
\caption{The relative deviation $\epsilon^{\rm mid}_{\rm X}(Q^2,t=2\,{\rm fm})$ for the midpoint estimates as a function of $Q^2$. Results for $X=E,4$ in red, $X=M$ in blue and $X=E,2$ in purple. Results for three different spatial volumes with $M_{\pi}L=4$ (diamonds), 5 (squares) and 6 (circles). Open symbols for the approximation with the excited-to-excited state contributions $c,\tilde{c}$ ignored, see main text.}
\label{fig:epsMid}
%\end{center}
\end{figure}

Figure \ref{fig:epsMid} shows the relative deviation for the midpoint estimates, i.e.
\bea
\epsilon^{\rm mid}_{\rm X}(Q^2,t)=\epsilon^{\rm eff}_{\rm X}(Q^2,t,t'=t/2)\,,
\eea
 as a function of $Q^2$ for $t=2\,$fm. Results are shown for three different spatial volumes with $M_{\pi}L=4$ (diamonds), 5 (squares) and 6 (circles). The results for a given volume show a smooth $Q^2$ dependence. A small FV effect is visible when we compare the results for $M_{\pi}L=4$ and $6$. However, it is much smaller than the anticipated precision of the LO results.

$\eV{4}$ is positive and rises monotonically to about $+5\%$ for $Q^2=0.25\,{\rm GeV}^2$. As discussed before, it vanishes for $Q^2=0$ as a result of the WI in \pref{VCCrelation3ptZERO}. The deviation $\eM$ for the magnetic form factor is negative and ranges between $-5\%$ and $-2\%$ for the momenta displayed in the figure. Here, in contrast to $\eV{4}$, the deviation increases for $Q^2$ getting smaller. 
Finally, the deviation $\eV{2}$ is close to $\eV{4}$, even though the difference between the two increases for small $Q^2$. Still, the difference is not pronounced enough to clearly favour one of the two ratios.

We emphasise that the low-momentum $N\pi$ contamination shown in fig.\ \ref{fig:epsMid} is the cumulative effect of many $N\pi$ states with different spatial momenta. For $M_{\pi}L=4$ we have taken into account all discrete momenta with $n_p$ up to 5, and this number rises to 12 for the larger volume with $M_{\pi}L=6$.

% Figure
% 
% End figure

Naively we expect the excited-to-excited state $N\pi$ contribution to be significantly smaller than the excited-to-ground state contribution. 
In terms of the coefficients we introduced this expectation says that the $b_\mu,\tilde{b}_\mu$ contributions are the dominant ones in eq.\ \pref{DefZmu} for $t'\approx t/2$. 
The reason is the additional suppression by an exponential factor $\exp(-\Delta E t/2)$ in the  $c_{\mu}, \tilde{c}_{\mu}$ contributions. 

% Figure
% 
\begin{figure}[p]
\begin{center}
$\epsilon^{\rm mid}_{\rm X,z}(Q^2,t=2\,{\rm fm})$\\
\includegraphics[scale=0.7]{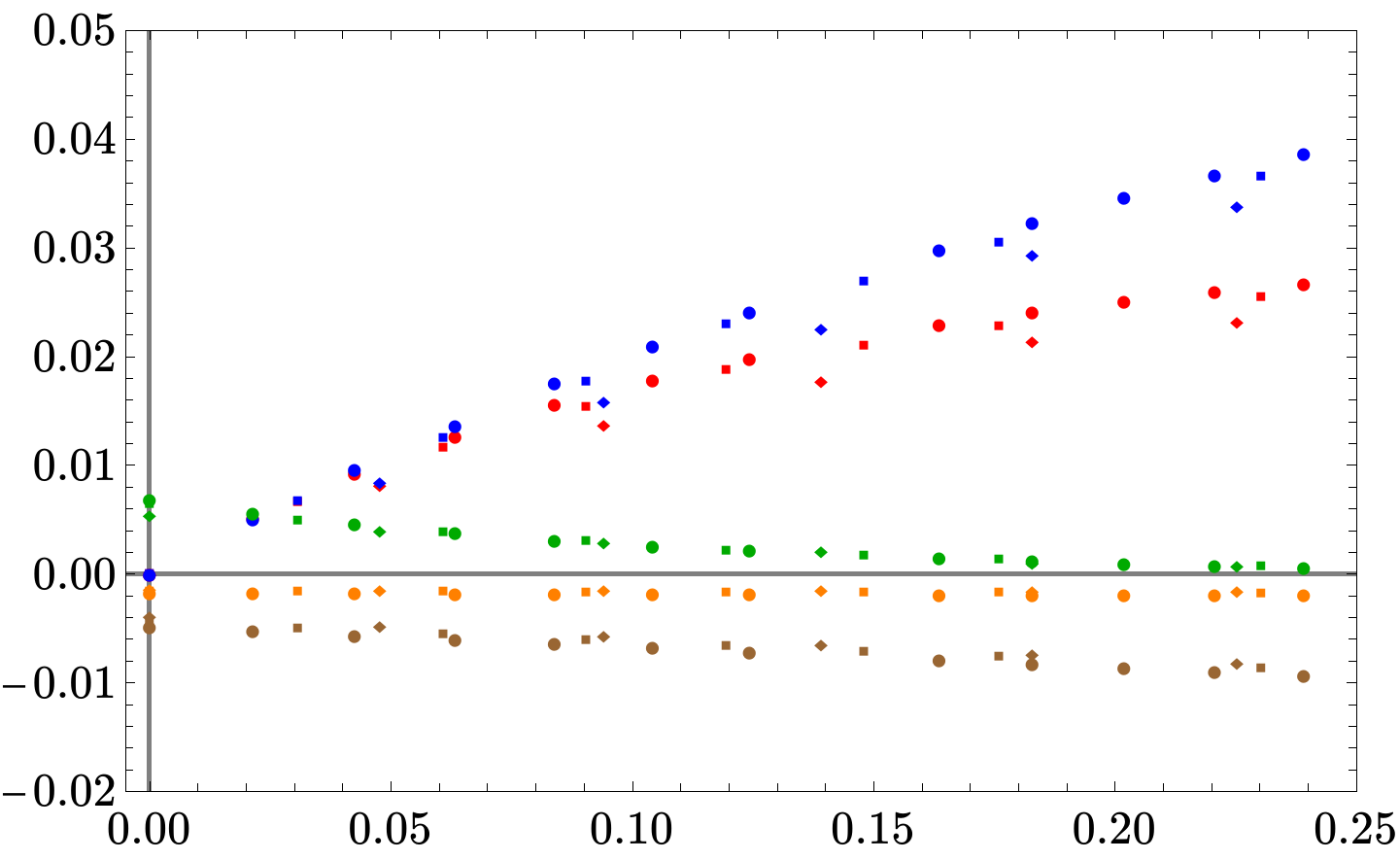}\\[2ex]
\includegraphics[scale=0.7]{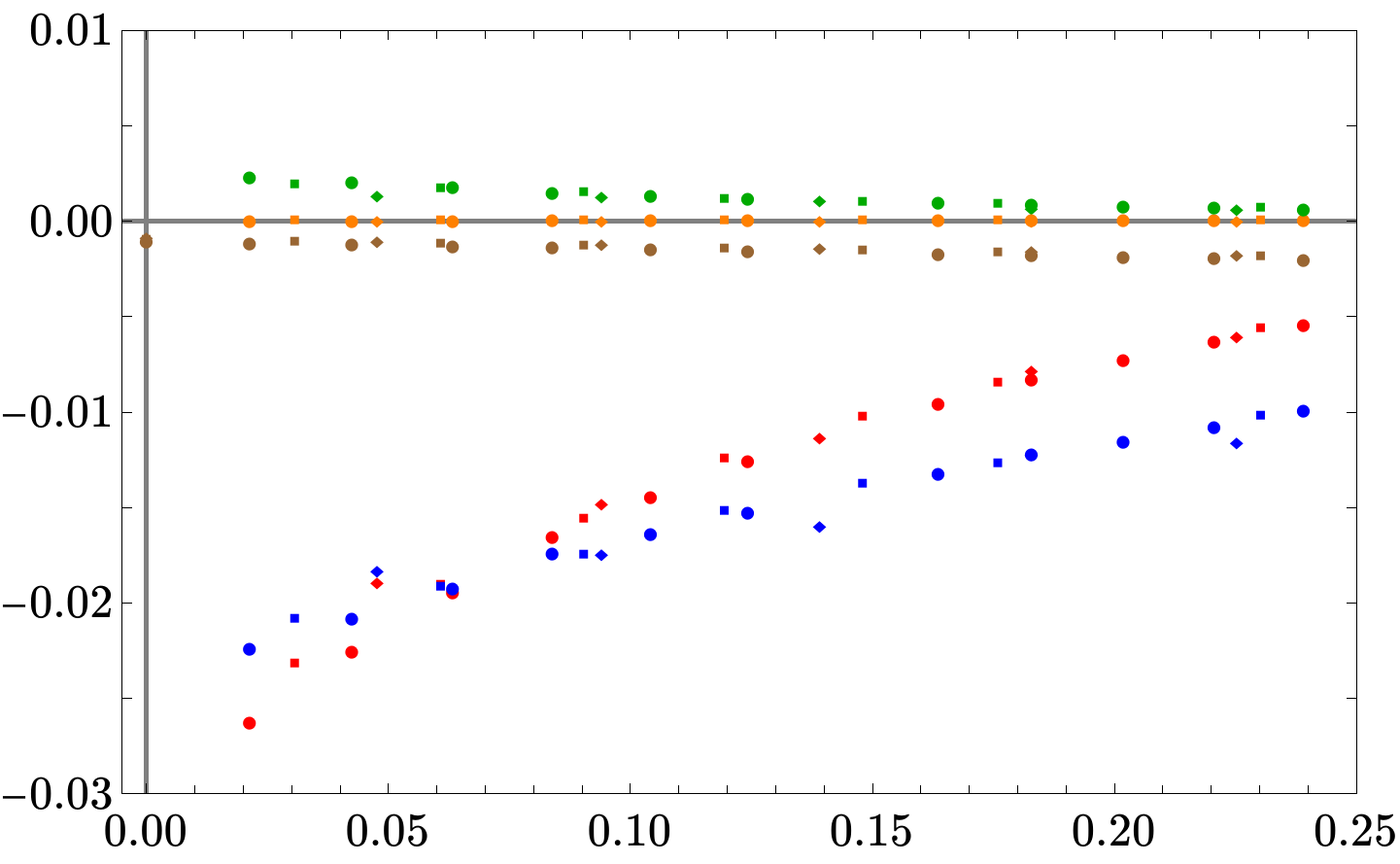}\\[2ex]
\includegraphics[scale=0.7]{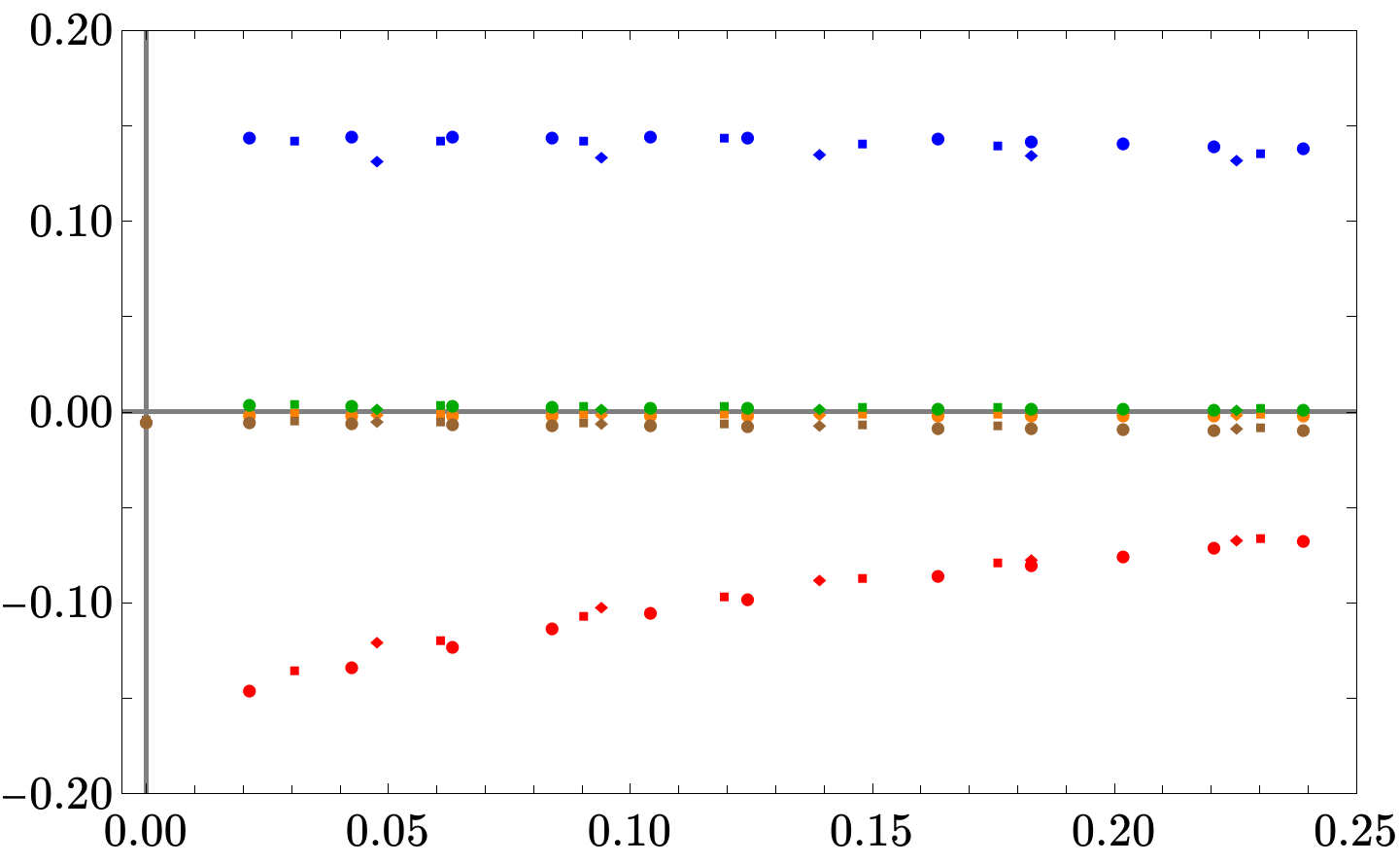}\\[1ex]
$Q^2/({\rm GeV})^2$\\[2ex]
\caption{The individual relative deviations $\epsilon^{\rm mid}_{\rm X,z}(Q^2,t=2\,{\rm fm})$ for the midpoint estimates as a function of $Q^2$. Results for contribution $z=b_{\mu}$ (red), $\tilde{b}_{\mu}$ (blue), $c_{\mu}$ (orange), $\tilde{c}_\mu$ (green) and $Y$ (brown) and for $X=E,4$ (upper panel), $X=M$ (middle) and $X=E,2$ (bottom). Results for three different spatial volumes with $M_{\pi}L=4$ (diamonds), 5 (squares) and 6 (circles).
Typically, the $b_{\mu}$ and $\tilde{b}_{\mu}$ contributions are larger than the other three.
}
\label{fig:epsMidcoeff}
\end{center}
\end{figure}
% End figure

Figure \ref{fig:epsMidcoeff} shows the individual contributions to the relative deviations, e.g.\ $\epsilon^{\rm mid}_{{\rm X},b}(Q^2,t)$ denotes the $b_\mu$ contribution (red symbols), and analogously for $\tilde{b}_\mu$ (blue), $c_\mu$ (orange) and $\tilde{c}_\mu$ (green). The $N\pi$ state contribution stemming from the 2-pt functions is shown by the brown symbols. 
Apparently, the $b_\mu$ and $\tilde{b}_\mu$ contributions are significantly larger than the other three. Note the relative sign between the $b_\mu$ and $\tilde{b}_\mu$ contributions in case of the $X=E,2$ (bottom panel), which is responsible for the sinh-like behaviour in the effective form factor $G^{\rm eff}_{{\rm E},2}$, seen in fig.\ \ref{fig:epsEff}.

The sum of all individual contributions in fig.\ \ref{fig:epsMidcoeff} gives the total results shown in fig.\ \ref{fig:epsMid}. Since the $c_\mu$ and $\tilde{c}_\mu$ contributions are small we can ignore them and still obtain a very good approximation for the total result. It is shown by the open symbols in fig.\ \ref{fig:epsMid}.

As stated before, the results shown so far are obtained with a finite number of $N\pi$ states in eq.\ \pref{DefY} and \pref{DefC3Npcontr}. The spatial momentum of the pion in the $N\pi$ state was restricted to  $|\vec{p}_n|\lesssim p_{\rm max}$ with $p_{\rm max}/\Lambda_{\chi}= 0.45$. We have checked that for $t=2$ fm these low-momentum $N\pi$ states essentially saturate the sums in \pref{DefY}, \pref{DefC3Npcontr} i.e.\ the contribution of the high-momentum $N\pi$ states is negligible. 

% Figure
% 
\begin{figure}[p]
\begin{center}
$\epsilon^{\rm mid}_{\rm E,4}(Q^2=Q^2_{n_q=5},t)$\\[0.4ex]
\includegraphics[scale=0.7]{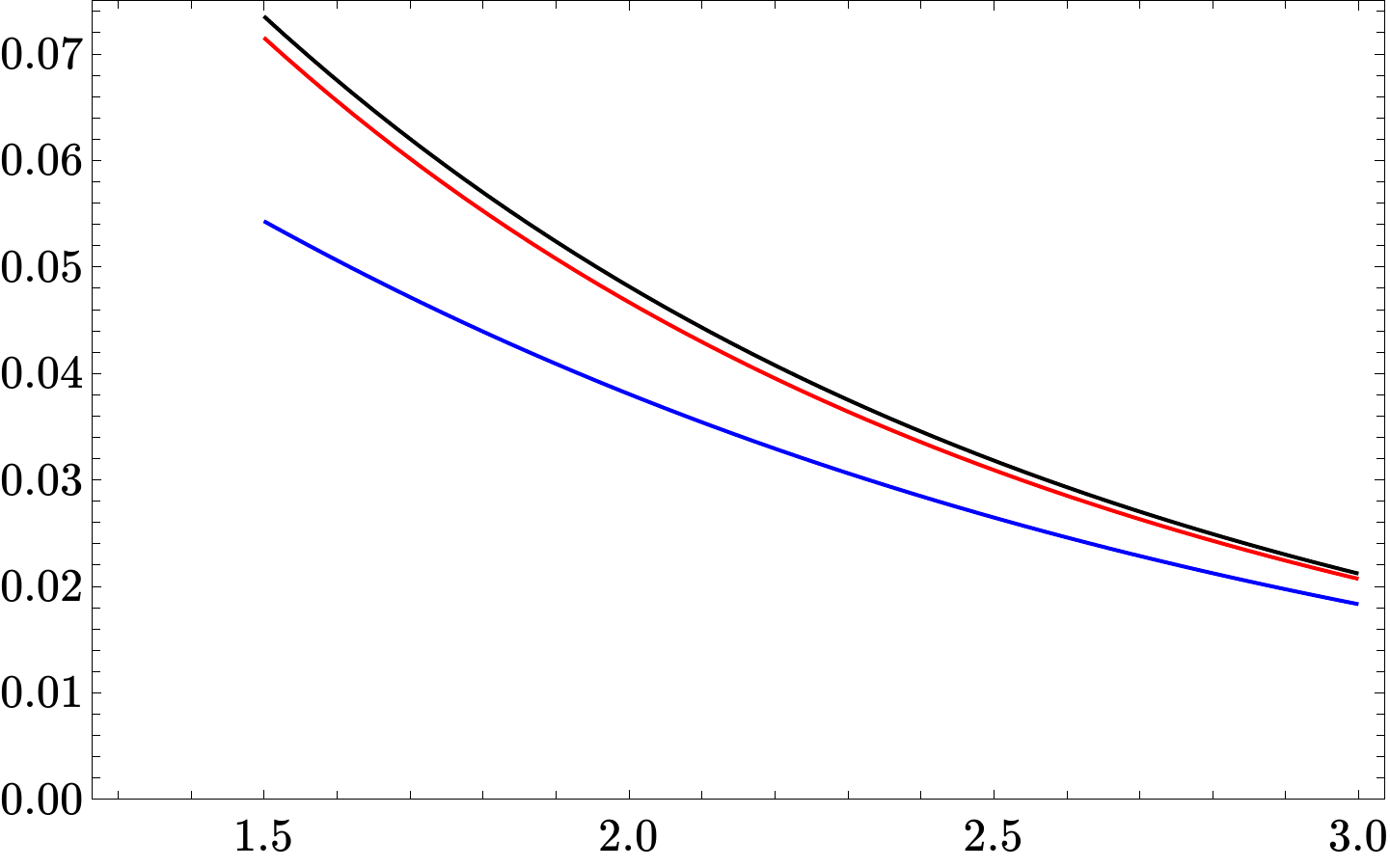}\\[0.4ex]
$t/{\rm fm}$\\[4ex]

$\epsilon^{\rm mid}_{\rm M}(Q^2=Q^2_{n_q=1},t)$\\
\includegraphics[scale=0.73]{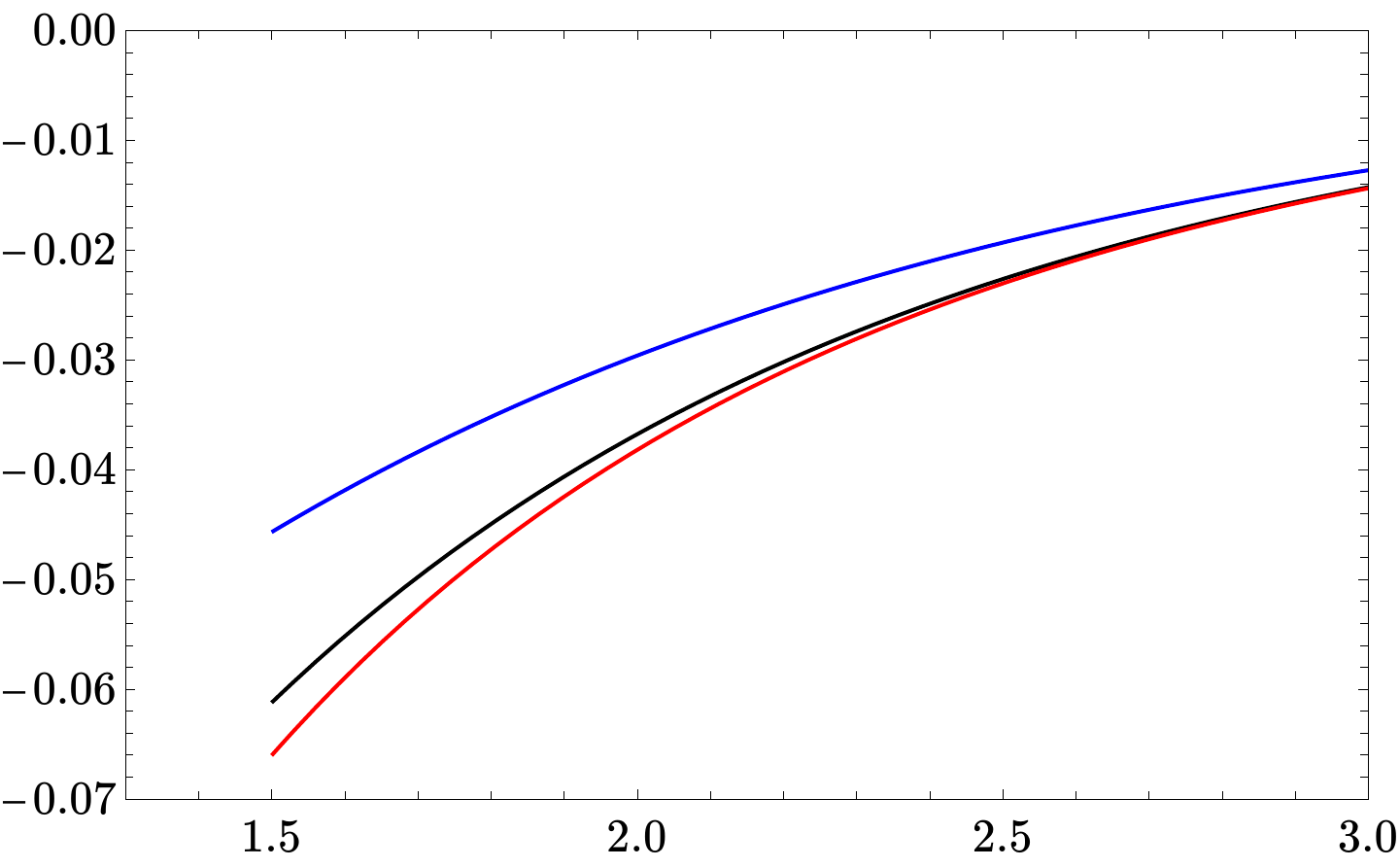}\\[0.4ex]
$t/{\rm fm}$\\[2ex]
\caption{The relative deviations $\epsilon^{\rm mid}_{\rm E,4}(Q^2,t)$ (top panel) and $\epsilon^{\rm mid}_{\rm M}(Q^2,t)$ (bottom) for the midpoint estimates as a function of the source-sink separation $t$. Results for $M_{\pi}L=4$ 
and $Q^2$ obtained with $n_q=5$ (top) and $n_q=1$ (bottom). 
}
\label{fig:epsDGX_of_t}
\end{center}
\end{figure}
% End figure

Two examples are shown in figure \ref{fig:epsDGX_of_t}. The lower panel shows the relative deviation $\epsilon^{\rm mid}_{\rm M}(Q^2,t)$ as a function of $t$ for $M_\pi L=4$ and the smallest accessible momentum transfer, i.e.\ with $n_q=1$. The black line corresponds to our canonical choice $p_{\rm max}/\Lambda_{\chi}= 0.45$. In addition, the results for two other momentum bounds are shown, a smaller one with $p_{\rm max}/\Lambda_{\chi}= 0.3$ (blue) and a larger one with $p_{\rm max}/\Lambda_{\chi}= 0.6$ (red). In terms of the integer $n_p$ these bounds correspond to $n_{p,{\rm max}}=2$ (blue), 5 (black) and 10 (red). For $t=2$ fm and larger the difference between the black and red curves is tiny and negligible, and this does not change if $p_{\rm max}$ is chosen even larger. However, a spread of about 25\% is seen between $p_{\rm max}/\Lambda_{\chi}= 0.3$ and 0.45 (for $t=2$ fm).

The upper panel in figure \ref{fig:epsDGX_of_t} shows the analogous result for $\epsilon^{\rm mid}_{\rm E,4}(Q^2,t)$, but for $n_q=5$, corresponding to $Q^2\approx 0.225 \,{\rm GeV}^2$. We find the same result, the low-momentum $N\pi$ states with $p_{\rm max}/\Lambda_{\chi}= 0.45$ essentially saturate the sum and capture the dominant part of the $N\pi$ excited-state contribution

\begin{figure}[t]
\begin{center}
$\GMmid(Q^2,t)/\GEmid(Q^2,t)$\\[0.4ex]
\includegraphics[scale=0.7]{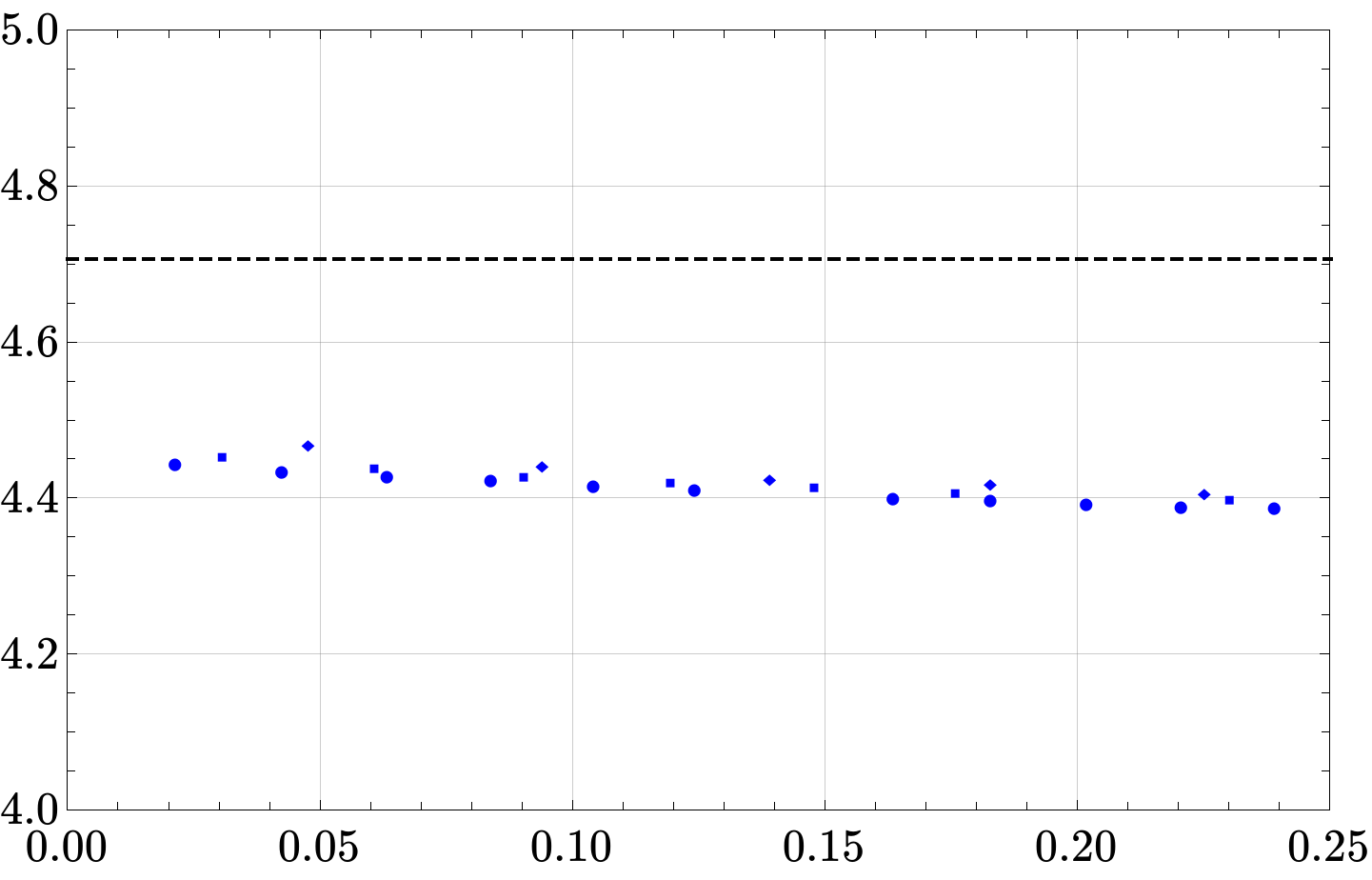}\\[0.4ex]
$Q^2/{\rm GeV}^2$\\[2ex]
\caption{\label{figGMmidOverGEmid} The ratio $\GMmid(Q^2,t)/\GEmid(Q^2,t)$ for $t=2$ fm and $M_{\pi}L=4$ (diamonds), 5 (squares) and 6 (circles). The dashed line shows the value $\mu_{p-n}=4.71$. 
}
\end{center}
\end{figure}

Finally, fig.\ \ref{figGMmidOverGEmid} shows the ratio $\GMmid(Q^2,t)/\GEmid(Q^2,t)$ as a function of $Q^2$, again for $t=2$ fm and various $M_\pi L$ values. To a very good approximation this ratio is constant, it varies by less than 2 percent over the range of $Q^2$ displayed in figure \ref{figGMmidOverGEmid}. This mild $Q^2$ dependence is anticipated since the slopes of $\eV{4}(Q^2,t)$ and $\eM(Q^2,t)$ are similar, see fig.\ \ref{fig:epsMid}, and essentially cancel in the ratio. However, this flat $Q^2$ behaviour should not be misinterpreted as the absence of the excited state contamination. The ratio is about 6\% below $\mu_{p-n}$, the value it assumes without the $N\pi$ contamination at vanishing momentum transfer.\footnote{Appendix F in Ref.\ \cite{Djukanovic:2021cgp} reports results of linear fits to lattice data for the ratio to extract the magnetic moment $\mu_{p-n}$. For almost all ensembles the extracted value is found well below  the experimental value, including an ensemble with physical pion mass. However, for this ensemble the maximal source-sink separation was $t\approx1.4$ fm.} Since $ \epsilon_\text{E}^\text{mid}(0,t)=0$ this underestimation stems dominantly from the $N\pi$ contamination in the magnetic form factor estimate $\GMmid(Q^2,t)$.

% Figure
% 
% End figure

\subsection{Impact on the charge radii}\label{ssect:impactonradii}

The $N\pi$ contamination in the form factor estimates is $Q^2$ dependent, and this affects the extraction of the charge radii. 
Estimates for the charge radii are obtained by eq.\ \pref{DefRadii}, with the midpoint estimates on the right hand side,
\begin{eqnarray}\label{DefRadiiMid}
\overline{r}^{2,{\rm mid}}_{\rm X}(t)\equiv -6 \frac{d}{d Q^2} \left(\frac{G^{\rm mid}_{\rm X}(Q^2,t)}{G^{\rm mid}_{\rm X}(0,t)}\right)\bigg|_{Q^2=0}\,.
\end{eqnarray}
In practice, the lattice form factor data are described by fitting a suitable  parameterisation of the $Q^2$ dependence, for example a dipole form or the $z$-expansion \cite{Hill:2010yb}. The charge radii are then obtained from these analytic forms. Here we can directly compute the derivative on the right hand side in \pref{DefRadii} to obtain an explicit expression for the midpoint estimates of the charge radii,
\begin{eqnarray}
\overline{r}^{2,{\rm mid}}_{\rm X}(t) \approx \overline{r}^{2}_{\rm X} -6 \epsilon_{\rm X}^{\prime,{\rm mid}}(0,t)\,.
\end{eqnarray}
The prime denotes the derivative $d/dQ^2$ of the relative deviation $\epsilon_{\rm X}^{{\rm mid}}$, and we have dropped terms quadratic or higher in $\epsilon_{\rm X}^{{\rm mid}}$. According to fig.\ \ref{fig:epsMid}  $\epsilon_{\rm X}^{\prime,{\rm mid}}(Q^2,t)$ is positive for both the electric and the magnetic form factors. Hence, both charge radii are underestimated by the midpoint estimates. 

It is straightforward to obtain the analytic expressions for $\epsilon_{\rm X}^{\prime,{\rm mid}}$, however, it is simpler and sufficient to obtain approximations directly from $\epsilon_{\rm X}^{{\rm mid}}$. For example, for $X={\rm E}$ we approximate $\epsilon_{\rm X}^{\prime,{\rm mid}}(0,t)\approx \epsilon_{\rm X}^{{\rm mid}}(Q^2_1,t)/Q^2_1$, where $Q^2_1$ is the smallest discrete momentum transfer displayed in fig.\ \ref{fig:epsMid}. With $\epsilon_{\rm E}^{{\rm mid}}(Q^2_1\approx 0.024{\rm GeV}^2,t)\approx 0.009$ we find that $\overline{r}^{{\rm mid}}_{\rm E}(t)$ is about 7\% smaller for $t=2$ fm  than the true charge radius $\overline{r}_{\rm E}$. Proceeding analogously for the magnetic form factor we find roughly $-4\%$ for the underestimation in case of $\overline{r}^{{\rm mid}}_{\rm M}(t)$.

%
%==============================
\section{Comparison with lattice data}
%==============================
%
 \subsection{Preliminaries}

In the last section we studied the $N\pi$ contamination for a source-sink separation of 2 fm, a rough lower bound for the ChPT results to be applicable with confidence. Present-day lattice calculations, however, are carried out at significantly smaller source-sink separations, the reason being the notorious signal-to-noise problem \cite{Parisi:1983ae,Lepage:1989hd} with exponentially growing statistical errors in the lattice data. 

Recent physical point simulations have been done with maximal source-sink separations $t_{\rm max} \approx 1.5$ fm. For instance, the ETMC collaboration reported results  \cite{Alexandrou:2018sjm} with $t_{\max}\approx 1.6$ fm for twisted mass fermions with a (charged) pion mass of about 140 MeV and a spatial lattice extent satisfying  $M_{\pi}L\approx 3.6$. While this volume may still be acceptable for sufficiently small FV effects it limits the accessible momentum transfers to only a few small values.
In contrast, recent simulations by the PACS collaboration \cite{Ishikawa:2018rew}  were performed with $M_\pi L\approx 6.0$ and close to physical pion mass, $M_{\pi} \approx 146$ MeV, admitting a smallest momentum transfer of about $Q^2 \approx 0.024\,{\rm GeV}^2$. However, the maximal source-sink simulation $t_{\rm max} \approx 1.3$ fm was even smaller than the one in the ETMC simulations.

Despite these shortcomings we compare the ChPT results to these lattice data, mainly to illustrate that the ChPT predictions for the $N\pi$ state contamination is qualitatively in agreement with what has been observed in lattice QCD data.\footnote{Other collaborations too have reported lattice form factor results obtained with (close to) physical pion masses. However, either the summation method \cite{Maiani:1987by,Capitani:2012gj} and/or multi-exponential fits have been used to obtain the form factor estimates (e.g.\ in \cite{Jang:2019jkn,Djukanovic:2021cgp,Park:2021ypf}) and our results are not applicable to those, or plateau estimates obtained at even smaller source-sink separations below 1 fm have been reported (e.g.\ in \cite{Hasan:2017wwt}), which seem way too small for a meaningful comparison with the ChPT results.}
Since the source-sink separation is significantly smaller than 2 fm we find a larger $N\pi$ contribution in the lattice estimates for the form factors. In addition, we may expect the contribution of other than low-momentum $N\pi$ states to be non-negligible. Thus, one should not be surprised that the ChPT results of the previous section do not fully account for the excited-state contribution observed in the lattice data.
   
\subsection{Electric and magnetic form factors}

% Figure
% 
\begin{figure}[p]
\begin{center}
$G_{\rm E,4}^{\rm plat}(Q^2,t)$\\[1ex]
\includegraphics[scale=0.7]{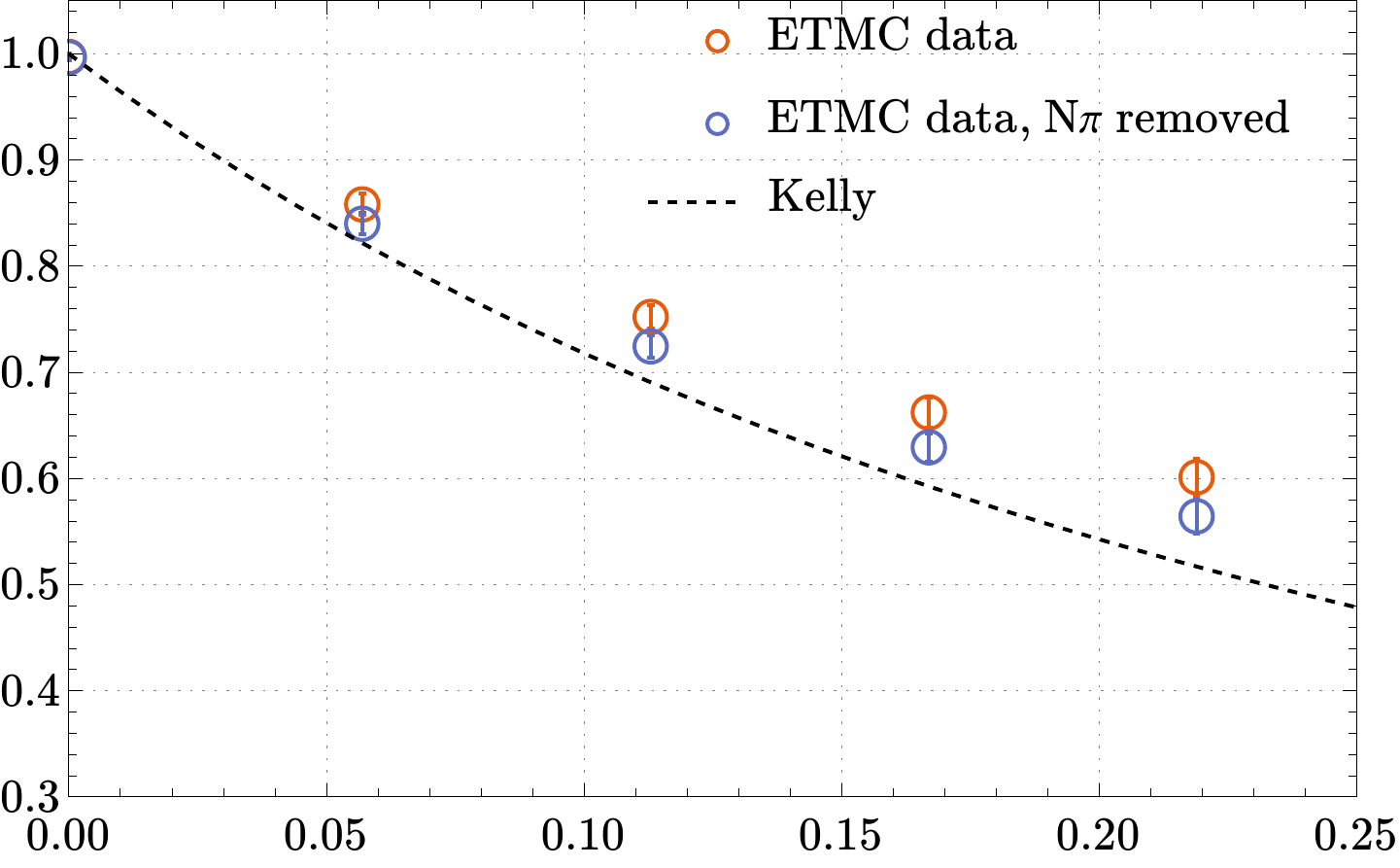}\\[0.4ex]
$Q^2/({\rm GeV})^2$\\[4ex]
$G_{\rm M}^{\rm plat}(Q^2,t)$\\[1ex]
\includegraphics[scale=0.7]{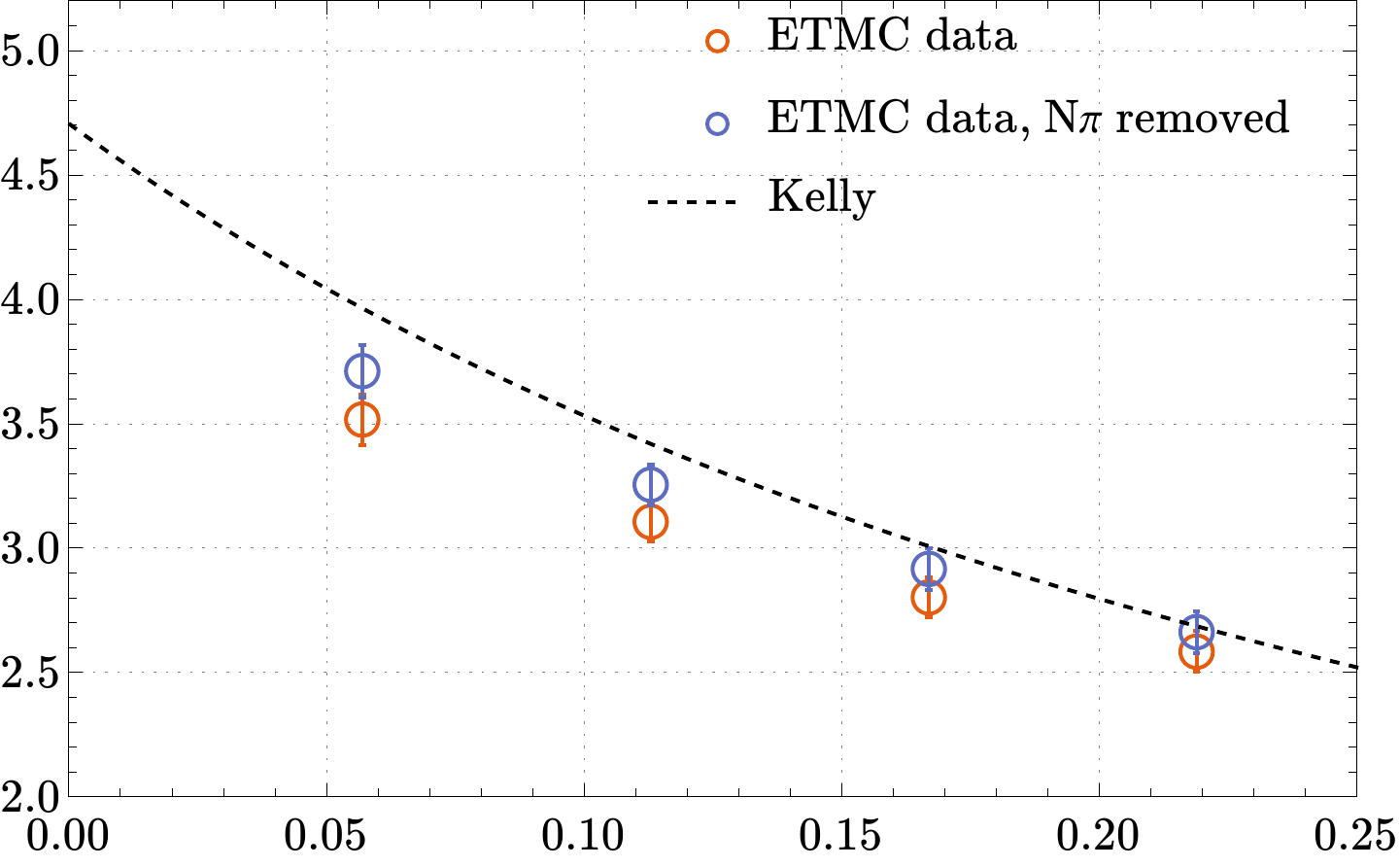}\\[0.4ex]
$Q^2/({\rm GeV})^2$\\[0.8ex]
\caption{\label{fig:ETMCGM1} ETMC data $G^{\rm plat}_{{\rm E},4}(Q^2,t)$ (top panel, orange symbols) and $G^{\rm plat}_{\rm M}(Q^2,t)$ (bottom panel, orange symbols) for $t=1.6$ fm and $Q^2$ smaller than $0.25\, {\rm GeV}^2$ \cite{Alexandrou:2018sjm}. The Kelly line (dashed) represents the experimental data \cite{Kelly:2004hm}. Removing the $N\pi$ contamination with eq.\ \pref{tindepofGPcorr} results in the corrected data points (blue symbols), see main text.
}
\end{center}
\end{figure}
% End figure

Figure \ref{fig:ETMCGM1} show the ETMC plateau estimates $G^{\rm plat}_{\rm E,4}(Q^2,t)$ (top panel) and $G^{\rm plat}_{\rm M}(Q^2,t)$ (bottom panel) for the electric and magnetic form factors, respectively (orange symbols).\footnote{See table VI in Ref.\ \cite{Alexandrou:2018sjm}.} The results were obtained with $N_f=2+1+1$ twisted mass clover-improved Wilson fermions at maximal twist and a lattice spacing $a\approx 0.08$  fm. The parameters relevant here are $M_{\pi}\approx139$ MeV, $M_N/M_{\pi}\approx 6.74$, $M_{\pi}L\approx 3.6$ and $t\approx 1.6$ fm. For more details we refer to \cite{Alexandrou:2018sjm}. 

The figure shows the data for the lowest momentum transfers below $Q^2=0.25\, {\rm GeV}^2$. Also shown is Kelly's fit (dashed lines) to the experimental data \cite{Kelly:2004hm}. In case of the electric form factor the lattice data lie above the Kelly line, overestimating the experimental results. The discrepancy increases for increasing momentum transfer. For the magnetic form factor the plateau estimates lie below the Kelly line, with the underestimation increasing for small $Q^2$. This agrees qualitatively with the ChPT results described in the last section for the impact of the $N\pi$ state contamination.

\begin{figure}[p]
\begin{center}
$\epsilon^{\rm plat}_{\rm E,4}(Q^2,t)$\\[0.6ex]
\includegraphics[scale=0.7]{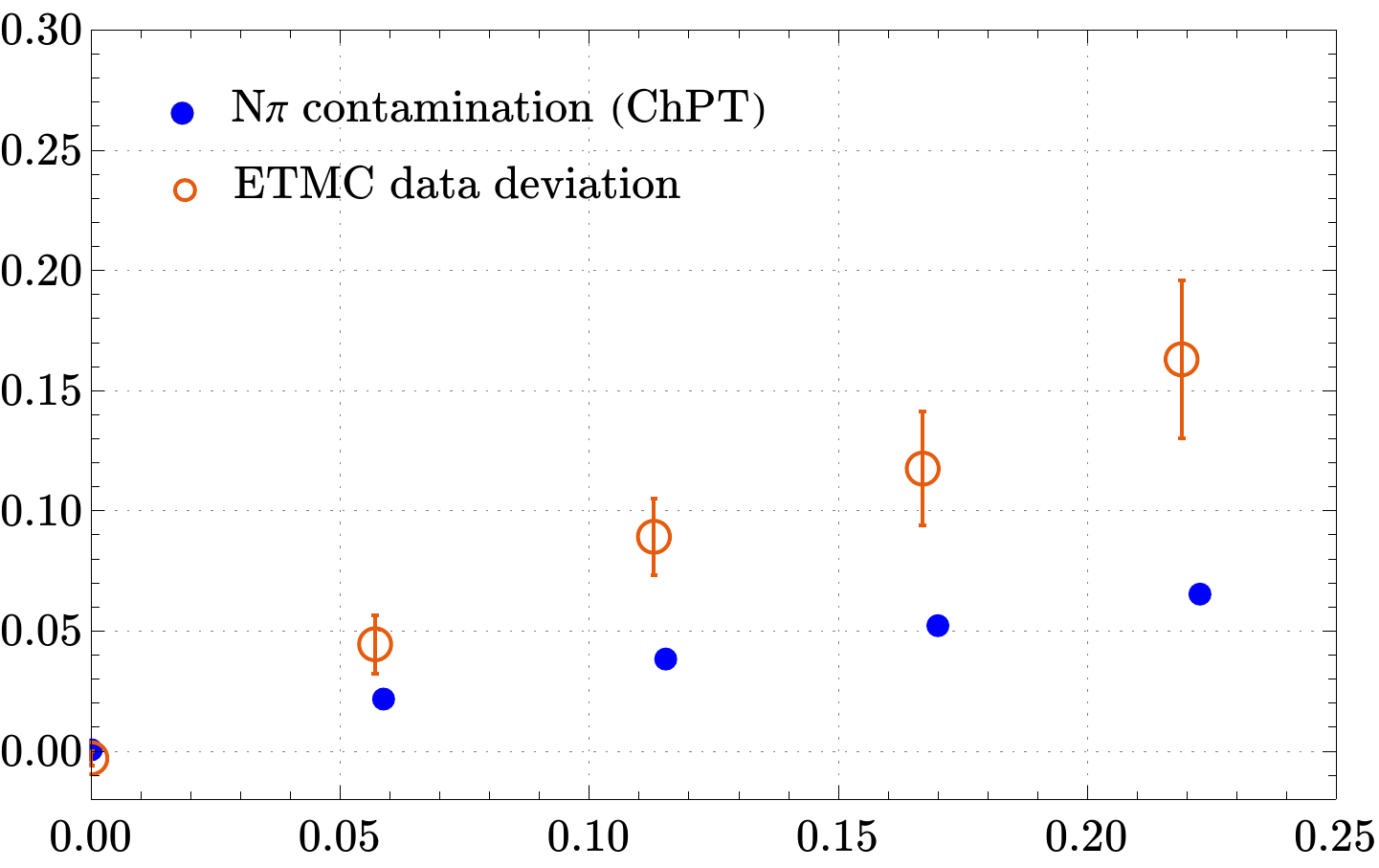}\\[0.4ex]
$Q^2/({\rm GeV})^2$\\[6ex]
$\epsilon^{\rm plat}_{\rm M}(Q^2,t)$\\[0.6ex]
\includegraphics[scale=0.7]{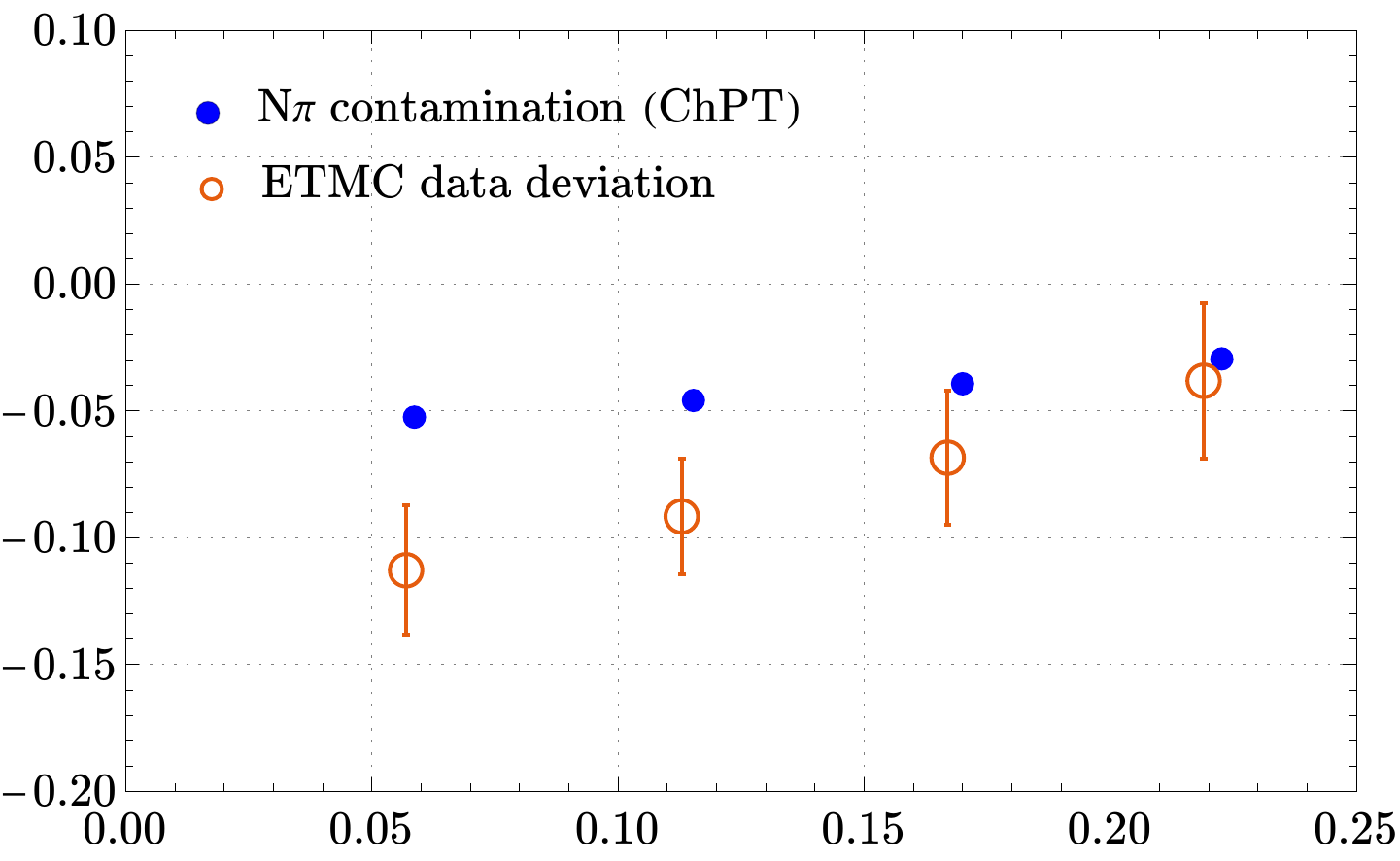}\\[0.4ex]
$Q^2/({\rm GeV})^2$\\[0.8ex]
\caption{\label{fig:ETMCGM1dev} The relative deviations $\epsilon^{\rm plat}_{\rm E,4}(Q^2,t)$ (top panel, orange symbols) and $\epsilon^{\rm plat}_{\rm M}(Q^2,t)$ (bottom panel, orange symbols)
of the ETMC data \cite{Alexandrou:2018sjm} from the Kelly line compared to the ChPT prediction for the deviation due to $N\pi$ states (blue). 
}
\end{center} 
\end{figure}

For a direct comparison, figure \ref{fig:ETMCGM1dev} shows the relative deviation of the lattice data from the Kelly line (orange symbols, with error bars) together with the ChPT results for $\eV{4}$ and $\eM$ (blue symbols). 
ChPT captures qualitatively the $Q^2$ dependence, but falls short roughly by a factor 2 for the largest (smallest) $Q^2$ in case of $\eV{4}$ ($\eM$) shown in the figure.

% Figure
% 
% End figure

There may be various sources for this discrepancy. We mentioned already that at $t=1.6$ fm we expect other than low-momentum $N\pi$ states to contribute to the total excited-state contamination. In addition, the ChPT results are obtained at LO only and may have a substantial higher order contribution in case of the magnetic form factor. Finally, the lattice date we compare with are obtained for a non-zero lattice spacing and may change when extrapolated to the continuum limit. With this in mind the fair agreement in fig.\ \ref{fig:ETMCGM1dev} 
is better than naively expected.

We can use the ChPT results in a slightly different but equivalent way. 
With the ChPT results for $\epsilon^{\rm plat}_{\rm X}(Q^2,t)$  we can ,,correct" the lattice data by analytically removing the LO $N\pi$-state contamination \cite{Bar:2018xyi}. For this we compute 
\begin{equation}\label{platestcorrected}
G^{\rm corr}_X(Q^2,t) \equiv \frac{G^{\rm plat}_X(Q^2,t) }{1+ \epsilon^{\rm plat}_{\rm X}(Q^2,t)}.
\end{equation}
In the numerator on the right hand side we take the lattice plateau estimates, while the denominator involves the ChPT results for the relative deviation. Provided higher order corrections and other than low-momentum $N\pi$-state contributions are small we conclude
\begin{equation}\label{tindepofGPcorr}
G^{\rm corr}_X(Q^2,t) \approx G_X(Q^2)\,,
\end{equation}
i.e.\ the corrected data should be essentially $t$ independent and equal to the true form factors. 
The corrected data is also shown in fig.\ \ref{fig:ETMCGM1} (blue symbols). The correction alleviates the discrepancy with the Kelly line, but a substantial deviation remains. 

% Figure
% 
\begin{figure}[p]
\begin{center}
$G_{\rm E,4}^{\rm plat}(Q^2,t)$\\[0.6ex]
\includegraphics[scale=0.7]{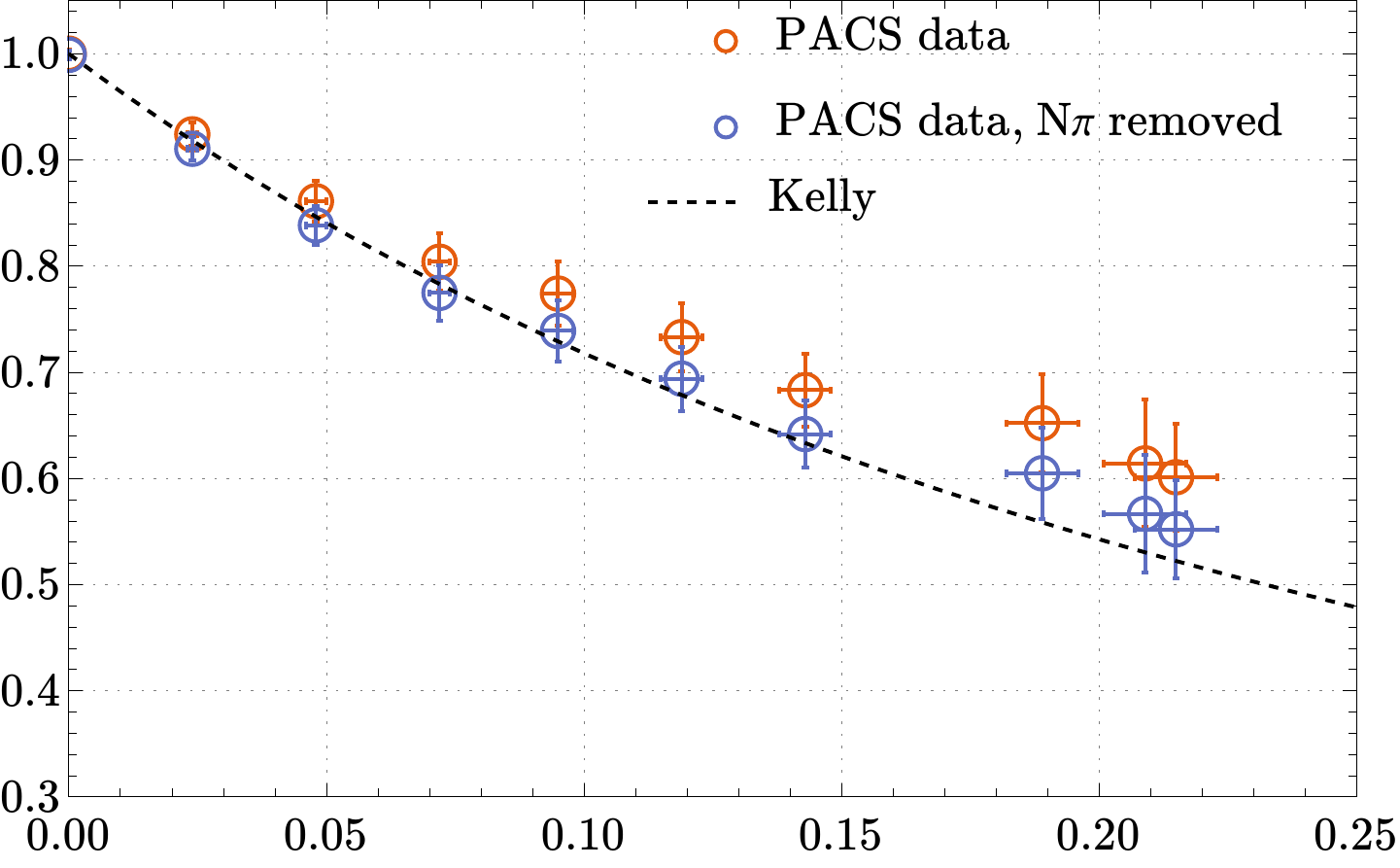}\\[0.6ex]
$Q^2/({\rm GeV})^2$\\[4ex]
$G_{\rm M}^{\rm plat}(Q^2,t)$\\[0.6ex]
\includegraphics[scale=0.7]{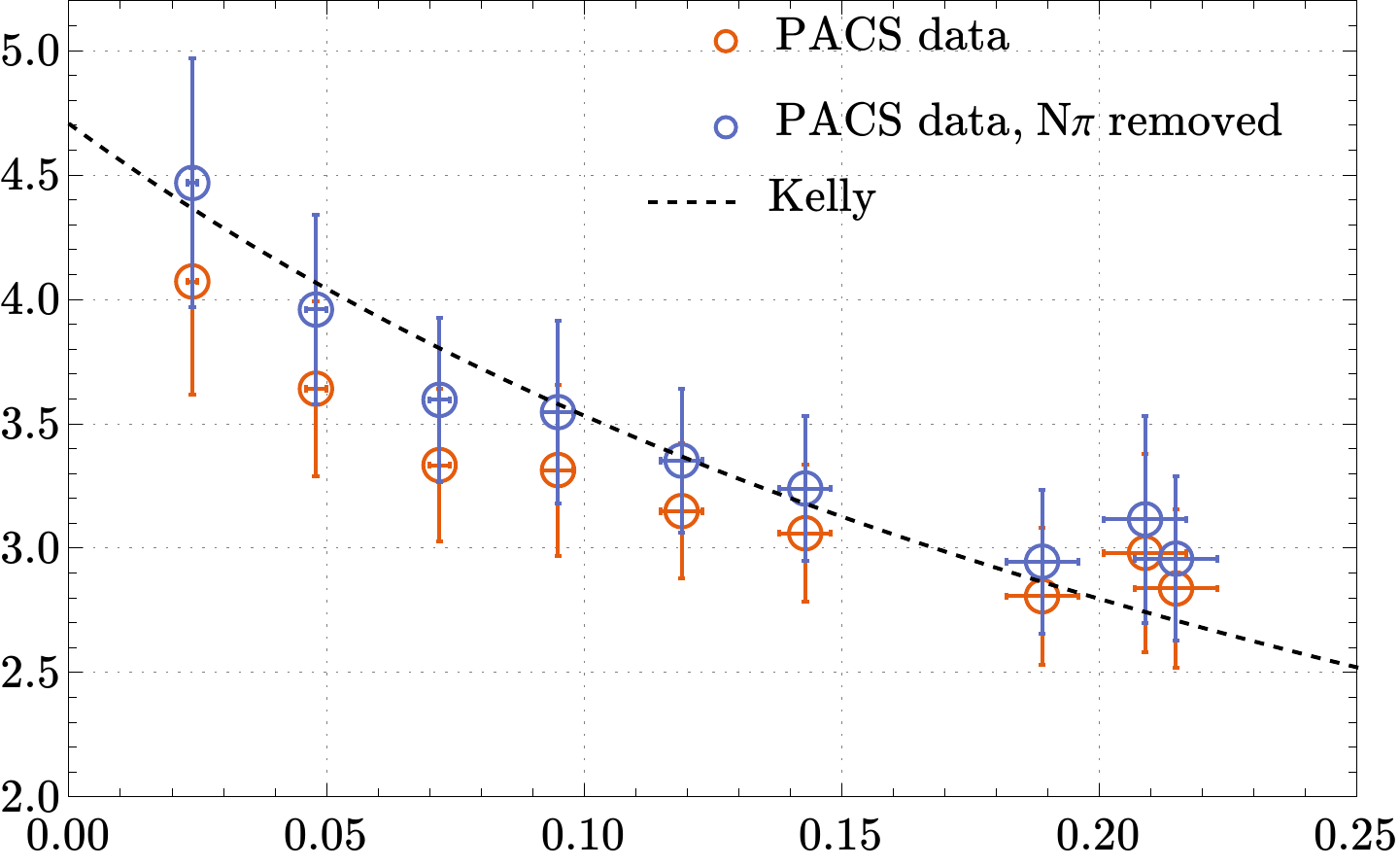}\\[0.4ex]
$Q^2/({\rm GeV})^2$\\[0.6ex]
\caption{\label{fig:PACSGM1} PACS data $G^{\rm plat}_{{\rm M}}(Q^2,t)$ (orange symbols) for $t=1.3$ fm and $Q^2$ smaller than $0.25\, {\rm GeV}^2$ \cite{Ishikawa:2018rew}. The Kelly line (dashed) represents the experimental data \cite{Kelly:2004hm}. Removing the $N\pi$ contamination with eq.\ \pref{tindepofGPcorr} results in the corrected data points (blue symbols), see main text.}
\end{center}
\end{figure}
% End figure

% Figure
% 
\begin{figure}[p]
\begin{center}
$\epsilon^{\rm plat}_{\rm E,4}(Q^2,t)$\\[1ex]
\includegraphics[scale=0.7]{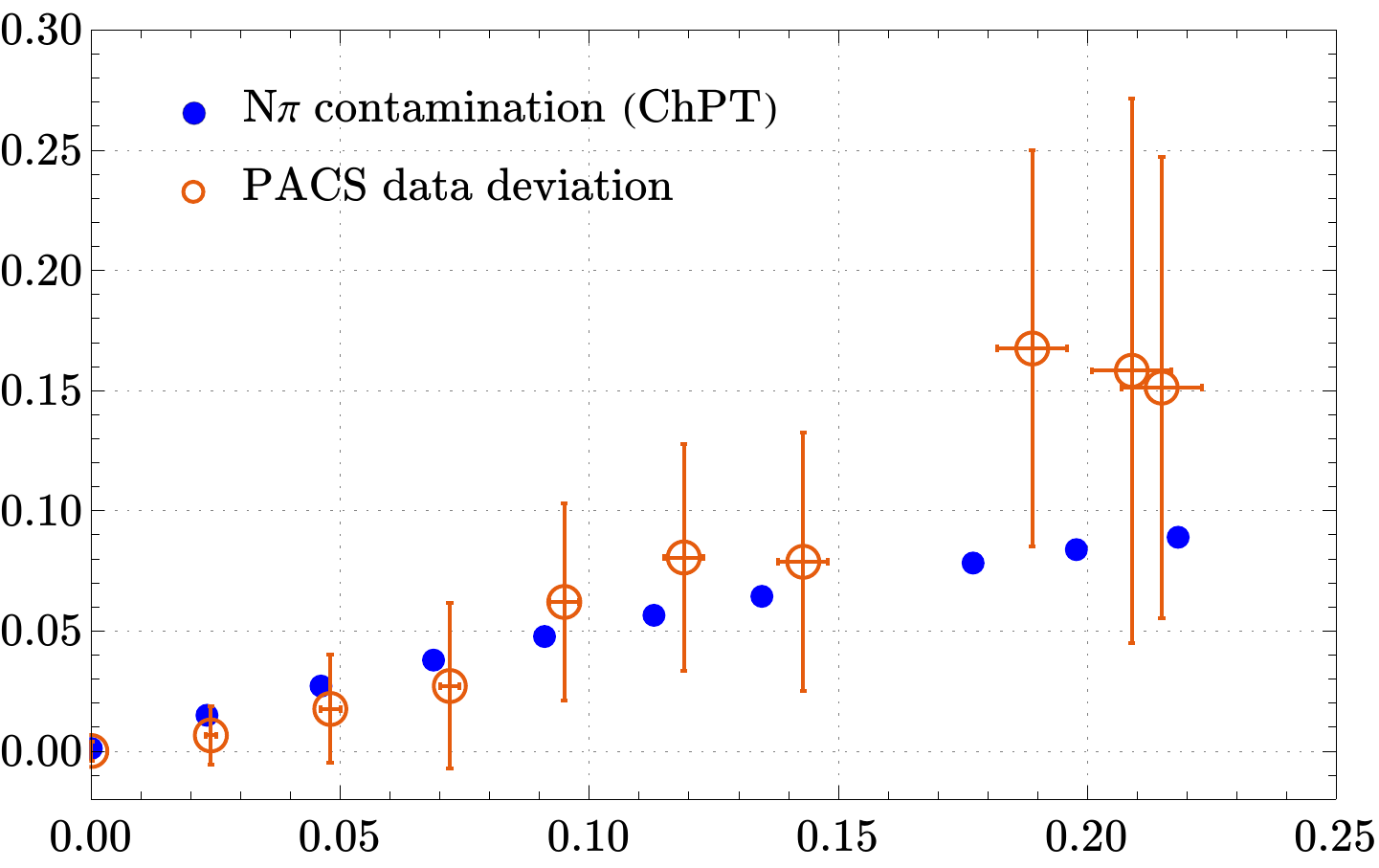}\\[0.6ex]
$Q^2/({\rm GeV})^2$\\[4ex]
$\epsilon^{\rm plat}_{\rm M}(Q^2,t)$\\[1ex]
\includegraphics[scale=0.7]{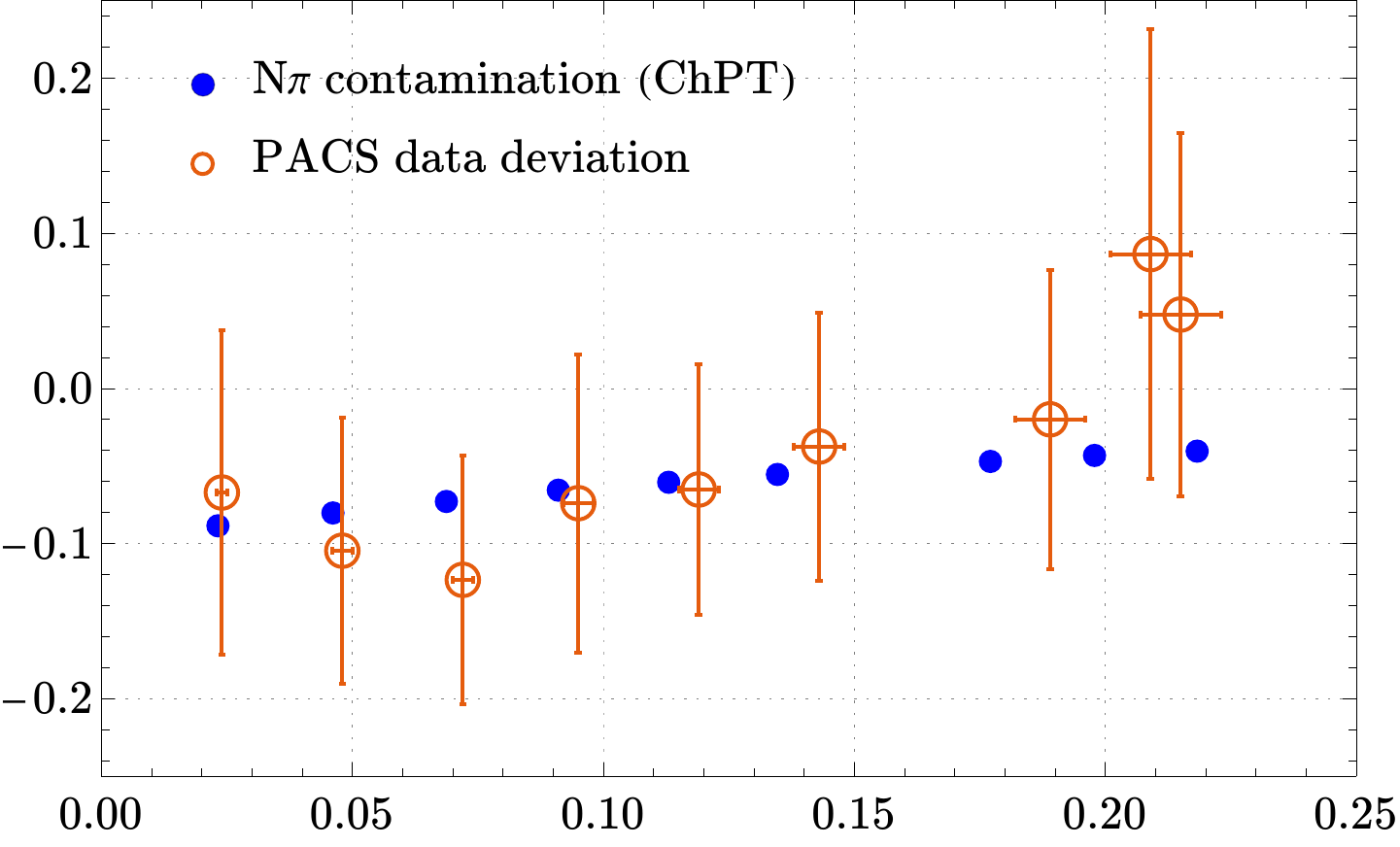}\\[0.6ex]
$Q^2/({\rm GeV})^2$\\[0.8ex]
\caption{\label{fig:PACSGM1dev} The relative deviations $\epsilon^{\rm plat}_{\rm M}(Q^2,t)$ 
of the ETMC data \cite{Alexandrou:2018sjm} from the Kelly line (orange symbols) compared to the ChPT prediction (blue).
}
\end{center}
\end{figure}
% End figure

Figs.\ \ref{fig:PACSGM1} and \ref{fig:PACSGM1dev} show the analogous data of the PACS collaboration, obtained with $N_f=2+1$ O($a$) improved Wilson fermions at $a\approx 0.085$ fm.\footnote{See table V in Ref.\ \cite{Ishikawa:2018rew}.} Due to the larger volume ($M_{\pi}L\approx 6.0$ with $M_{\pi} \approx146$ MeV and $M_N/M_{\pi}\approx 6.56$) smaller momentum transfers are accessible. Less favourable are the even smaller source-sink separation $t\approx 1.3$ fm and the larger statistical errors.

We observe the same qualitative features as in the ETMC data. The plateau estimates (red symbols) deviate from the Kelly line the same way: The electric form factor is overestimated, the magnetic one underestimated. Also shown are the corrected form factors (blue symbols). The improvement looks better to the eye, but it is statistically not significant because of the large statistical errors. 

\subsection{Comment on the spatial correlator for the electric form factor}

As discussed before, the electric form factor is accessible either with the time-like or a spatial component of the vector current in the 3-pt function. 
In a recent publication \cite{Jang:2019jkn} the PNDME collaboration studied and compared both ways.
The lattice data were obtained in a mixed-action setup with 
highly improved staggered sea quarks (HISQ) and clover improved Wilson valence quarks.
Among the ensembles are two with a (close-to) physical pion mass and a reasonably large spatial volumes satisfying $M_{\pi}L\approx3.9$ and $3.7$, respectively.  
For more details on the lattice ensembles see Refs.\ \cite{Jang:2019jkn,Gupta:2018qil}. 

The data displayed in figs.\ 24 and 25 of \cite{Jang:2019jkn} show indeed the sinh-like behaviour that we found for the spatial correlator, in contrast to the time-like correlator which displays the cosh-like behaviour (recall top and bottom panel in fig.\ \ref{fig:epsEff}). Moreover, the excited-state contamination in the spatial correlator is found to be an order of magnitude larger -- for the smallest non-vanishing momentum transfer -- as compared to those from the time-like correlator. This too agrees qualitatively with the O($M_N$) enhancement of the $N\pi$ contamination in the ratio with the spatial correlator, cf.\ eq.\ \pref{EnhancedNRExp}.

The main obstacle for a more qualitative comparison is the analysis strategy used in \cite{Jang:2019jkn}. Instead of simple plateau or midpoint estimates, multi-exponential fit ansaetze were used to extract the form factors with simultaneous fits to the 2-pt and 3-pt function data. Proceeding this way the excited-state contamination we are interested in here  is at least partially removed from the form factor estimators. 

\section{Conclusions}

The ChPT results presented here provide an understanding for the anticipated impact of $N\pi$ excited states in lattice computations of the nucleon electromagnetic form factors. ChPT predicts an overestimation of the electric form factor by the plateau or midpoint estimates, and this overestimation gets larger for increasing momentum transfer. For the magnetic form factor we find the opposite impact, it is underestimated and the smaller the momentum transfer, the larger the misestimation. The size of this effect is about $\pm 5$\% for source-sink separations of 2 fm and small momentum transfers smaller than $0.25\,{\rm GeV}^2$. If percent level accuracy is the goal for lattice calculations this source of systematic error is not negligible and needs to be taken care of.

The effect is larger for source-sink separations that are presently accessible with standard simulation methods, for $t\approx1.6\,{\rm fm}$ by roughly a factor two. However, we stress once again that at this $t$ we probably have pushed the ChPT results beyond their range of applicability. At such rather small source-sink separations we expect the excited-state contamination not to be dominated by low-momentum $N\pi$ states alone. Additional excited states are expected to contribute as well, with potentially large modifications to the results given here. Comparison with more lattice data at larger $t$ is needed to elucidate this issue.

A subleading excited-state contamination is caused by three-particle $N\pi\pi$ states. In general their contribution stems from 2-loop diagrams with two propagators for the two pions, and these are expected to be very small \cite{Bar:2018wco}. However, there are also some 1-loop diagrams responsible for an $N\pi\pi$ contribution, for instance diagrams n) to p) in fig.\ \ref{fig:Npidiags3pt}. We have checked this contribution and found it, not surprisingly, to be of the same size as the $N\pi$ contribution captured by $\tilde{c}$, to which these diagrams also contribute. Since the $\tilde{c}$ contribution is small and negligible we expect the same for the entire $N\pi\pi$ contribution, even though this has to be confirmed by a complete calculation.

A common way to cope with the excited-state contamination is the use of multi-exponential fit ansaetze in the analysis of lattice data. In a recent publication \cite{Park:2021ypf} a variety of different ansaetze have been used and compared with each other, including ansaetze that fix the energy gaps to the lowest $N\pi$ excited state. In most cases these ansaetze work equally well as the ones that leave the energy gaps free parameters to be determined by the fits. However, in some cases the results for the form factors differ significantly depending on the fit ansatz chosen. This exemplifies the need for physical understanding of the excited-state contamination in lattice data. Judging the quality of the fits by the $\chi^2$ values alone might be misleading and insufficient for percent level calculations of the form factors.

Finally, in addition to understanding lattice data at larger source-sink separations seem mandatory for obtaining precise lattice results with controlled errors. To overcome the notorious signal-to-noise problem new simulation techniques are required. The idea proposed in Refs.\ \cite{Ce:2016idq,Ce:2016ajy} seems promising but remains to be tested in lattice calculations of nucleon correlation functions.

\vspace{4ex}
\noindent {\bf Acknowledgments}
O.B.\ acknowledges useful discussions with Rajan Gupta and his sharing of manuscript \cite{Park:2021ypf} before publication.
This work was supported by the German Research Foundation (DFG), Grant ID BA 3494/2-1.
\vspace{3ex}

\noindent{\large \bf Appendix}

%========================
\begin{appendix}
%========================
\section{ChPT setup}

\subsection{Summary of the Feynman rules}\label{appFeynmanRules}

We employ the covariant formulation of baryon ChPT \cite{Gasser:1987rb,Becher:1999he}.
To LO the chiral effective Lagrangian consists of two parts, ${\cal L}_{\rm eff}={\cal L}_{N\pi}^{(1)} + {\cal L}_{\pi\pi}^{(2)}$. Expanding this Lagrangian in powers of pion fields and keeping interaction terms with one pion field only we obtain
\begin{eqnarray}
\label{Leff}
{\cal L}_{\rm eff} &=& \overline{\Psi} \Big(\gamma_{\mu}\partial_{\mu} +M_N \Big)\Psi +\frac{1}{2}\pi^a \Big(- \partial_{\mu}\partial_{\mu} + M_{\pi}^2 \Big)\pi^a + \frac{ig_A}{2f}\overline{\Psi}\gamma_{\mu}\gamma_5\sigma^a \Psi \, \partial_{\mu} \pi^a\,.
\end{eqnarray}
The  nucleon fields $\Psi=(p,n)^T$ and $\overline{\Psi}=(\overline{p},\overline{n})$ 
contain the proton and the neutron fields $p$ and $n$.  $M_{\pi}$ denotes the pion mass, while $M_N$, $g_A$ and $f$ are the chiral limit values of the nucleon mass, the axial charge and the pion decay constant. The interaction term in \pref{Leff} leads to the well-known nucleon-pion interaction vertex proportional to the axial charge. A factor $i$ appears here because we work in Euclidean space-time. 

From the terms quadratic in the fields one reads off the nucleon and pion propagators. For our calculations the time-momentum representation seems most convenient. In that representation the pion propagator reads
\begin{eqnarray}
G^{ab}(x,y)& = &  \delta^{ab}L^{-3}\sum_{\vec{p}} \frac{1}{2 \Epip} e^{i\vec{p}(\vec{x}-\vec{y})} e^{-\Epip |x_0 - y_0|}\,,\label{scalprop}
\end{eqnarray}
with  the pion energy given by $\Epip =\sqrt{\vec{p}^2 +M_{\pi}^2}$. The nucleon propagator $S^{ab}_{\alpha\beta}(x,y)$ is given by
\begin{eqnarray}
S_{\alpha\beta}^{ab}(x,y)& = &  \delta^{ab} L^{-3}\sum_{\vec{p}} \frac{Z_{p,\alpha\beta}^{\pm}}{2E_N} e^{i\vec{p}(\vec{x}-\vec{y})} e^{-E_N |x_0 - y_0|}\,.
\end{eqnarray} 
$a,b$ and $\alpha,\beta$ refer to the isospin and Dirac indices, respectively. The factor $Z^{\pm}_{\vec{p}}$ in the nucleon propagator (spinor indices suppressed) is defined as
\begin{equation}
Z_{\vec{p}}^{\pm}=-i\vec{p}\cdot\vec{\gamma} \pm E_N \gamma_4+M_N\,, 
\end{equation}
where the $+$ ($-$) sign applies to $x_0 > y_0$ ($x_0 < y_0$), and the nucleon energy is given by $E_{N,\vec{p}}=\sqrt{|\vec{p}|^2 +M_N^2}$. 
The sum in both propagators runs over the discrete spatial momenta that are compatible with periodic boundary conditions imposed on the finite spatial volume, i.e.\ 
$\vec{p}=2\pi \vec{n}/L$  with $\vec{n}$ having integer-valued components.

The expressions for the nucleon interpolating fields in ChPT have been derived in Ref.\ \cite{Wein:2011ix}. To LO and up to one power in pion fields one finds
\begin{eqnarray}\label{Neffexp}
N(x)& = & \tilde{\alpha} \left(\Psi(x) + \frac{i}{2f} \pi^a(x)\sigma^a \gamma_5\Psi(x)\right)\,,\\
\overline{N}(0) & = & \tilde{\beta}^* \left(\overline{\Psi}(0) + \frac{i}{2f}\overline{\Psi}(0)\gamma_5\sigma^a\pi^a(0) \right)
\end{eqnarray}
These are the effective fields for the standard nucleon interpolating fields composed of three quarks without derivatives \cite{Ioffe:1981kw,Espriu:1983hu}. The interpolating fields not necessarily need to be point-like, but can also be constructed from `smeared' quark fields. These operators map to the same chiral expressions provided the smearing procedure is compatible with chiral symmetry. In addition, the `smearing radius' needs to be small compared to the Compton wavelength of the pion. In that case smeared interpolating fields are mapped onto point like fields in ChPT just like their pointlike counterparts at the quark level  \cite{Bar:2013ora,Bar:2015zwa}. The expressions differ only by the LECs $\tilde{\alpha},\tilde{\beta}$. If the same interpolating fields are used at both source and sink we find $\tilde{\alpha}=\tilde{\beta}$. 

Finally, for the computation of the 3-pt function we need the expression for the vector current.
It is obtained from the effective Lagrangian in the presence of an external source field for the vector current \cite{Gasser:1987rb}. The source field enters via covariant derivatives acting on the nucleon and pion fields, guaranteeing invariance of the Lagrangian under local chiral transformations.

The LO Lagrangian ${\cal L}_{\rm eff}={\cal L}_{N\pi}^{(1)} + {\cal L}_{\pi\pi}^{(2)}$ leads to the following expression for the vector current:
\begin{eqnarray}
V_{\mu}^{a,(1)} & = & \overline{\Psi}\gamma_{\mu} \sigma^a\Psi -\frac{g_A}{f}\epsilon^{abc} \pi^b \overline{\Psi}\gamma_{\mu}\gamma_5\sigma^c\Psi - 2i \epsilon^{abc}\partial_{\mu} \pi^b \pi^c\,.\label{DefVectorCurrent}
\end{eqnarray}
The first two terms on the right hand side stem from ${\cal L}_{N\pi}^{(1)}$, the remaining one from ${\cal L}_{\pi\pi}^{(2)}$. The same expression has already been used in Refs.\ \cite{Bar:2016uoj} for the computation of the nucleon vector charge $g_V$. 

For the computation of the electromagnetic form factors we also need the contribution to the vector current from the nucleon-pion Lagrangian with chiral dimension two,
\begin{equation}\label{L2Lag}
{\cal L}^{(2)}_{N\pi}=c_6\frac{M_N}{f^2} \overline{\Psi}\sigma_{\mu\nu}f^+_{\mu\nu}\Psi\,  +\,\ldots \,,
\end{equation}
with the usual tensor $\sigma_{\mu\nu} = [\gamma_{\mu},\gamma_{\nu}]/2$ and $f^+_{\mu\nu}$ being the field strength tensor formed with the external source field \cite{Gasser:1987rb}. We have omitted terms, represented by the ellipses, that are not relevant in the following. $c_6$ is a LEC that can be related to the magnetic moments of the proton and neutron, see next subsection. The term in \pref{L2Lag} leads to the contribution 
\begin{eqnarray}
V_{\mu}^{a,(2)} & = & 4 c_6\frac{M_N}{f^2} \left(\partial_{\alpha}\overline{\Psi} \sigma_{\mu\alpha}\sigma^a\Psi+\overline{\Psi} \sigma_{\mu\alpha}\sigma^a\partial_{\alpha}\Psi \right)\label{DefVectorCurrent2}
\end{eqnarray}
in the vector current (dropping all terms with more than one pion field). For the calculation in this paper the current $V_{\mu}^a=V_{\mu}^{a,(1)}+V_{\mu}^{a,(2)}$ was used.

\subsection{Single nucleon results}\label{app:SNresults}
With the Feynman rules in the last subsection we obtain the diagram in fig.\ \ref{fig:diagsSN} for the leading SN contribution in the 3-pt function. The analogous diagram for the 2-pt function is essentially the same but without the vector current insertion. The calculation of these diagrams is trivial, and forming the generalised ratio \pref{DefRatio} with these results we obtain
\begin{eqnarray}
{\rm Re}\,R^N_{V^3_{4}} (\vec{q},t,t')& = & \sqrt{\frac{\ENq+M_N}{2\ENq}} \left(1- \frac{\ENq -M_N}{2 M_N} \tilde c_6\right)      \,,\label{RSN4}\\
{\rm Re}\,R^N_{V^3_{i}}(\vec{q},t,t') & = & \epsilon_{ij3}q_j \frac{1}{\sqrt{2\ENq(\ENq +M_N)}}  \left(1+ \tilde c_6\right)\,,\label{ReRSNk}\\
{\rm Im}\,R^N_{V^3_{i}}(\vec{q},t,t') & = & q_i \frac{1}{\sqrt{2\ENq(\ENq +M_N)}}  \left(1- \frac{\ENq -M_N}{2 M_N} \tilde c_6\right)\,,\label{ImRSNk}
\end{eqnarray}
with the short hand notation $\tilde c_6 = 8 M_N^2c_6/f^2$. By construction the time dependence cancels for the SN results, hence the rhs correspond to the asymptotic values \pref{AsympValuePi4} - \pref{AsympValueImPik} of the ratios. Comparing with these expressions we read off the leading SN results for the electromagnetic form factors,
\begin{eqnarray}\label{FFSNLO}
G_{\rm E}(Q^2)&  =  & 1- \frac{\ENq -M_N}{2 M_N} \tilde c_6\,,\quad
G_{\rm M}(Q^2) \, = \, 1+\tilde c_6\,.
\end{eqnarray}
The result $G_{\rm E}(0)=1$ agrees with the constraint set by the WI, as expected. The leading result for the magnetic form factor is independent of the momentum transfer. We fix the LEC $\tilde c_6$ by setting \pref{FFSNLO} to the phenomenological value at $Q^2=0$,
\bea\label{GMdeltamu}
\GM(0)=\mu_p - \mu_n\,,
\eea
the difference of the magnetic moments of proton and neutron. The experimental value for this difference is 4.706 \cite{Zyla:2020zbs}. 

Note that the $\tilde c_6$ contribution is rather large. With the contribution in \pref{DefVectorCurrent} only we find $\GM(0)=1$, which is a crude approximation and significantly underestimates \pref{GMdeltamu}. The SN results enter the $N\pi$ contamination in the denominator of the coefficients in \pref{DefZmu}. Therefore, without the contribution \pref{DefVectorCurrent2} for the vector current we would overestimate the ratios by roughly a factor of five.

Even though $c_6$ enters $\GE$ as well its impact is much smaller. In fact, performing the NR expansion in \pref{FFSNLO} we find $\GE(Q^2)= 1+{\rm O}(1/M_N^2)$. Since we only work to ${\rm O}(1/M_N)$ our results for the electric form factor in section \ref{ssect:Npicontribution} do not depend on $c_6$.

%===========
\section{Results for the correction coefficients}\label{app:corrcoeff}
%===========
The correction coefficients $B_{k}^{\rm corr}(\vec{q},\vec{p}),\tilde{B}_{k}^{\rm corr}(\vec{q},\vec{p}),C_{k}^{\rm corr}(\vec{q},\vec{p})$ and $\tilde C_{k}^{\rm corr}(\vec{q},\vec{p})$ are introduced in section \ref{ssect:Npicontribution} and listed here. 
The results are rather lengthy, so we have split them in $O(g_A)$ and $O(g_A^2)$ contributions, with the complete result being the sum of the two contributions. For the spatial indices $k$ and $m$ applies $ m, k = 1, 2$ and $ m \neq k$.

For the real part of the spatial components we find\\[0.3cm]
$O(g_A)$:
\begin{align}
	B^{re,\text{corr}}_k &= -\frac{ B^{re,\infty}_k}{2g_A}
	\\
	\tilde{B}^{re,\text{corr}}_k &= \frac{4 g_A (p_m+q_m)}{q_m}
	\\
	C^{re,\text{corr}}_k &= -\frac{ C^{re,\infty}_k}{g_A} - \frac{g_A}{2}
						\left( \frac{p_m}{q_m} \frac{\qsq}{\Epip^2} - \frac{2 p_k}{q_m} \frac{( p_k q_m - p_m q_k )}{\Epip^2} \right)
	\\
	\tilde{C}^{re,\text{corr}}_k &= -\frac{ \tilde{C}^{re,\infty}_k}{2g_A}						
\end{align}
$O(g_A^2)$:
\begin{align}
	B^{re,\text{corr}}_k &= \frac{M_\pi^2}{2\Epip^2} B^{\infty, re}_k  -
		\frac{2 g_A^2}{\Epip^2 q_m} \left( \Epip^2 p_m - \psq q_m + p_3 ( p_m q_3 - p_3 q_m) \right)
	\\
	\tilde{B}^{re,\text{corr}}_k &= -\frac{1}{2} \left( \frac{\psq + 2 \pq}{\Epip^2} + 1 \right)
		\tilde{B}^{re,\infty}_k 
		\\ \nonumber
		&+ \frac{2 g_A^2}{\Epip^2 q_m} \left(\psq q_m - p_3 (p_m q_3 - p_3 q_m) - \Epip^2 (3 p_m + 4 q_m)\right)
	\\
	C^{re,\text{corr}}_k &= -\frac{\psq + \pq}{\Epip^2} C^{\infty, re}_k  +
		\frac{g_A^2}{\Epip^2 q_m} \left( 2 p_m \pq + \qsq p_m - \psq q_m \right)
	\\
	\tilde{C}^{re,\text{corr}}_k &= -\frac{\tilde{C}^{re,\infty}_k}{2 \Epip^2 \Epipmq} 
		\left(2 \Epip ( \psq - \pq ) + M^2_\pi ( \Epip - \Epipmq ) \right)
\end{align}
For the {real} part of the {time} component we get\\[0.3cm]
$O(g_A)$:

\begin{align}
	B^{re,\text{corr}}_4 &= -\frac{ B^{re,\infty}_4}{2g_A}
	\\
	\tilde{B}^{re,\text{corr}}_4 &= - \frac{\tilde{B}^{re,\infty}_4}{2g_A} 
		+ 2g_A \left(\frac{\qsq + \pq}{\Epippq^2} - \frac{\pq}{\Epip^2} \right)
\end{align}
\begin{align}
	C^{re,\text{corr}}_4 &= -\frac{C^{re,\infty}_4}{g_A}
		+ \frac{g_A}{2\Epip^2} \pq
	\\
	\tilde{C}^{re,\text{corr}}_4 &= -\frac{\tilde{C}^{re,\infty}_4}{2 g_A}
		- \frac{g_A}{\Epip^2 \Epipmq} (\Epip + \Epipmq) \psq
\end{align}
$O(g_A^2)$:
\begin{align}
	B^{re,\text{corr}}_4 &= \frac{ B^{re,\infty}_4}{2} 
		-g_A^2 \frac{\psq}{\Epip^2} \frac{\left(4 \psq + \pq \right)}{\Epip^2}
		\\
		&+ \frac{g_A^2}{\Epip^2 \Epipmq^2} \left( 2(\psq - \pq)(2 \psq - \qsq) - \Epipmq^2 \ \pq \right)
	\nonumber \\
	\tilde{B}^{re,\text{corr}}_4 &= \left( \frac{1}{2} - \frac{\psq + 2\pq}{\Epip^2} \right) \tilde{B}^{re,\infty}_4
		+ \frac{g_A^2}{\Epip^4} \pq \left( \Epip^2 + \psq  \right) 
		\\
		&+ \frac{2g_A^2}{\Epip^2 \Epippq^2} \left( \pq ( 4 \pq + 3 \qsq ) - \qsq ( M_\pi^2 + \Epip^2)\right)
	\nonumber \\
	C^{re,\text{corr}}_4 &= \frac{M^2_\pi}{\Epip^2} C^{re,\infty}_4
		- \frac{g_A^2 M^2_\pi}{\Epip^4} \pq
	\\
	\tilde{C}^{re,\text{corr}}_4 &= \frac{\tilde{C}^{re,\infty}_4}{2} 
		\left(\frac{\qsq - \psq}{\Epip \Epipmq} - \frac{\psq}{\Epip^2} \right) \\
		&+ \frac{g_A^2}{\Epip^2 \Epipmq^2} \left( \psq (\Epip + \Epipmq)^2 - \pq (\Epip^2 - \Epipmq^2) \right) \nonumber
\end{align}
For the {imaginary} part of the {spatial} components we get\\[0.3cm]
$O(g_A)$:
\begin{align}
	B^{im,\text{corr}}_k &= -\frac{ B^{im,\infty}_k}{2g_A}
	\\
	\tilde{B}^{im,\text{corr}}_k &= -\frac{\tilde{B}^{\infty, im}_k}{2 g_A} 
		+ 4 g_A \left( 1 - \frac{(2p_k + q_k)}{q_k}\frac{(\qsq + \pq)}{\Epippq^2} \right)
	\\
	C^{im,\text{corr}}_k &= -\frac{C^{im,\infty}_k}{g_A}
		+ \frac{g_A}{2} \frac{p_k}{q_k} \frac{(\qsq + 2\pq)}{\Epip^2}
	\\
	\tilde{C}^{im,\text{corr}}_k &= -\frac{\tilde{C}^{im,\infty}_k}{2 g_A}
		+ 2 g_A \frac{(2 p_k - q_k)}{q_k} \frac{\psq}{\Epip \Epipmq}
\end{align}
$O(g_A^2)$:
\begin{align}
	B^{im,\text{corr}}_k &= \frac{M^2_\pi}{2\Epip^2} B^{im,\infty}_k
		+ 2 g_A^2 \frac{p_k}{q_k} + 2 g_A^2 \frac{(p_k + 2 q_k)}{q_k} \frac{\psq}{\Epip^2}
	\\[0.3cm]
	\tilde{B}^{im,\text{corr}}_k &= \frac{1}{2} \left( 1 - \frac{\psq + 2\pq}{\Epip^2} \right) \tilde{B}^{\infty, im}_k + 2 g_A^2 \left( \frac{p_k}{q_k} \left( 1 + \frac{\psq}{\Epip^2} \right) + 2 \frac{\psq}{\Epip^2} \right)
		\\
		&- 8 g_A^2 \left( 1 - \frac{(2p_k + q_k)}{q_k} \frac{( \qsq + \pq )}{\Epippq^2} \right)
	\nonumber\\[0.3cm]
	C^{im,\text{corr}}_k &= -\frac{\psq + \pq}{\Epip^2}C^{im,\infty}_k
		- g_A^2 \left( \frac{( 2 p_k + q_k )}{q_k} \frac{\psq}{\Epip^2} + \frac{p_k}{q_k} \frac{(\qsq + 2 \pq)}{\Epip^2} \right)
	\\[0.3cm]
	\tilde{C}^{im,\text{corr}}_k &= \frac{M^2_\pi}{2 \Epip^2} \tilde{C}^{im,\infty}_k
		- \frac{2g_A^2}{\Epip \Epipmq} \frac{2p_k - q_k}{q_k} \frac{\left( M^2_\pi ( \psq + \pq ) + 2 ( \psq \qsq - (\pq)^2)\right)}{\Epipmq^2}
\end{align}
 
\end{appendix}

%======================

%======================

\end{document}